%% file: nmaddox.tex
\documentclass[useAMS]{mn2e}

\usepackage{graphicx}
\usepackage{amssymb}
\usepackage{lscape}

\title{Luminous $K$-band Selected Quasars from UKIDSS}

\author[N. Maddox, P.C. Hewett, S.J. Warren, S.M. Croom]{Natasha
  Maddox\thanks{nmaddox@aip.de}$^{1,2}$, Paul C. Hewett$^1$,
  S.J. Warren$^3$, S.M. Croom$^4$
\vspace*{6pt}\\
$^1$Institute of Astronomy, University of Cambridge, Madingley Road,
Cambridge CB3 0HA, UK \\
$^2$Astrophysikalisches Institut Potsdam, An der Sternwarte 16, 14482 Potsdam,
Germany \\
$^3$Astrophysics Group, Imperial College London, Blackett Laboratory,
Prince Consort Road, London, SW7 2AZ, UK\\
$^4$School of Physics, University of Sydney, NSW 2006, Australia\\
}

\begin{document}


\maketitle

\begin{abstract}
  The largest $K$-band flux-limited sample of luminous quasars to date
  has been constructed from the UKIDSS Large Area Survey Early Data
  Release, covering an effective area of 12.8~deg$^2$. Exploiting the
  $K$-band excess of all quasars with respect to foreground stars,
  including quasars experiencing dust reddening and objects with
  non-standard SEDs, a list of targets suitable for spectroscopic
  follow-up observations with the AAOmega multi-object spectrograph is
  constructed, resulting in more than 200 confirmed AGN. KX-selection
  successfully identifies as quasar candidates objects that are
  excluded from the SDSS quasar selection algorithm due to their
  colours being consistent with the stellar locus in optical colour
  space (with the space density of the excluded objects agreeing well
  with results from existing completeness analyses). Nearly half of
  the KX-selected quasars with $K\le17.0$ at z $<3$ are too faint in
  the $i$-band to have been targeted by the SDSS quasar selection
  algorithm, revealing a large population of quasars with red $i-K$
  colours. The majority of these objects have significant amounts of
  host galaxy light contributing to their $K$-band magnitudes,
  consistent with previous predictions. The remaining objects are
  morphologically stellar and have colours consistent with quasars
  experiencing SMC-type reddening with $0.10<$~E($B-V$)~$<0.25$. The
  $i-K$ colour distribution indicates that $<10$~per~cent of the
  quasar population is missing from this $K$-band selected sample due
  to dust reddening, and comparisons with simulations strongly favour
  an obscured fraction of $<20$~per~cent. Photometric redshifts and
  classifications are computed for the candidates that were not
  observed spectroscopically. For the extended objects whose colours
  are consistent with those of a reddened quasar, models of galaxy
  surface brightness profiles appropriate for each object are used to
  eliminate the possibility of the presence of a nuclear source bright
  enough for inclusion in a $K\le17.0$ quasar sample. The
  effectiveness of near-infrared colour selection of quasars has been
  demonstrated by this modest-sized sample, and it will only become
  more apparent as the amount of available data increases.

\end{abstract}

\begin{keywords}
quasars:general--surveys--infrared:general
\end{keywords}


\section{Introduction}

For many years, the most popular method for selecting quasar
candidates has exploited the fact that their spectral energy
distribution (SED) produces an excess of flux at short wavelengths
compared to the blackbody radiation from stars. This UV-excess (UVX)
method has been very efficient at finding `typical' blue quasars at
redshifts less than z~$=2$. However, it introduces an unknown bias
into the samples, excluding any quasar that has an SED that departs
from the assumed standard power-law. Multicolour selection with
relaxed morphology constraints, as employed by the Sloan Digital Sky
Survey (SDSS) (Richards et al. 2002), reduces this bias, but does not
eliminate it. However, even the multicolour optical selection breaks down
at redshifts near z~$\sim 2.7$, as the stellar locus intersects the
region of colour-space occupied by quasars, resulting in low
completeness. In addition to excluding some quasars with
atypical SEDs, optical selection is incapable of selecting quasars
experiencing even a small amount of dust reddening, as passbands at
the shorter wavelengths are significantly affected by dust
attenuation.

A complete census of quasar activity over cosmic time is essential for
understanding galaxy formation and evolution, as evidenced by the
intimate relation between properties of the central black hole and
those of its host galaxy. However, due to the large range in quasar
SED properties, no single observational selection encompasses the
entire diverse population. Although it is fairly well established that
at least some of the differences in SED appearances are due to dust
reddening, there is no consensus regarding whether these differences
can be explained in the context of a unified model or whether they are
a consequence of evolution throughout a quasar's lifetime.

The existence of heavily dust obscured quasars is not contested, but
the debate regarding the fraction of quasars missing from existing
samples due to dust extinction is ongoing, with results ranging from
$\sim$15 per cent (Richards et al. 2003) to 60 per cent and higher
(see, for example, Glikman et al. 2007, White et al. 2003). A key
question is whether the obscured fraction is luminosity dependent, the
answer to which would provide important constraints within which any
successful scenario would be required to fit.

In order to determine the fraction of optically obscured luminous
quasars, a successful experiment must cover large areas to interesting
magnitude limits, at wavelengths for which dust extinction is not
severe. Deep X-ray studies are capable of detecting all but the most
heavily obscured quasars, but their small areas are insufficient for
finding the relatively rare, luminous quasars in statistically
significant numbers. The Two Micron All Sky Survey (2MASS, Skrutskie
et al.~2006) essentially covers the entire sky, but the bright
magnitude limits restrict the detected population to all but the most
luminous quasars, or those very nearby. The new generation of
near-infrared (NIR) wide field surveys provides the opportunity to
define a sample of quasars, to interesting flux limits, that is
sensitive to objects experiencing moderate amounts of dust reddening.

\begin{figure}
\resizebox{\hsize}{!}{\includegraphics{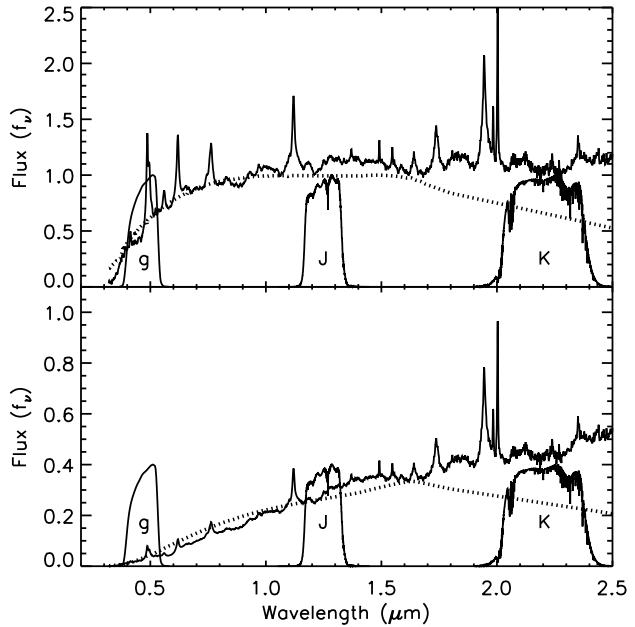}}
\caption{Comparison of stellar and quasar SEDs, with the SDSS $g$,
  UKIDSS $J$ and $K$ passbands overplotted (Hewett et
  al. 2006). Top: The solid line is a composite quasar spectrum at
  z $=3$, the dotted line is an early K-star, showing a clear $K$-band
  excess. Bottom: The solid line is the same composite quasar, with restframe
  extinction E$(B-V)=0.3$, the dotted line is an early M-star. A
  significant $K$-band excess is still observed. The figure is adapted
  from Warren et al. (2000).}
\label{fig:seds}
\end{figure}

Analogous to the UV-excess seen at shorter wavelengths, there is an
equivalent excess in the $K$-band (KX) for quasars compared to stars
(Warren, Hewett \& Foltz 2000), as illustrated in Fig.
\ref{fig:seds}.  Both top and bottom panels of Fig. \ref{fig:seds}
show quasars overplotted with stars of similar $g-J$ colours, but in
each case the quasar exhibits a clear $K$-band excess, even when the
quasar is reddened by dust. This illustrates the power of the
KX-method as it is insensitive to colour changes due to reddening.

There have been two studies to date employing KX-selection of quasars;
the first by Croom, Warren \& Glazebrook (2001) and the second by
Sharp et al. (2002). Although pioneering in nature, the limited
areal coverage, 48\,arcmin$^2$ and 0.7\,deg$^2$, respectively,
restricts the effectiveness in constraining the fraction of reddened
quasars.

The current work combines optical imaging and spectroscopy from the SDSS, new
NIR imaging from the UKIDSS project, and optical multi-object
spectroscopy using the new AAOmega spectrograph to create a
well-defined sample of luminous quasars, flux-limited in the $K$-band
to $K\le17.0$, covering 14~deg$^2$. Close to 100 per cent
identification success is required for the entire sample of quasar
candidates in order to place meaningful restrictions on the fraction
of optically obscured quasars. For the objects observed
spectroscopically, secure classifications are generally
straightforward to achieve. For the candidates without spectra, a
combination of photometric identifications/redshifts, determined using
$ugrizYJHK$ photometry, and morphological properties in both the
optical and NIR, are employed to provide classifications of high
reliability.

The experiment is deliberately set up to undertake as comprehensive a
survey as possible with no regard to the question of efficiency. The
rationale for such an approach is that it has not been previously
demonstrated that an efficient survey that was also highly effective
at selecting quasars could be undertaken using this particular
combination of optical and NIR data.

The outline of the paper is as follows. Section 2 describes the
selection criteria used to create the sample to be observed
spectroscopically. The spectroscopic data is presented in Section 3,
along with a description of the photometric redshift determination
used to classify candidates not observed spectroscopically. Section 4
presents the composition of the new $K$-band selected sample of
objects, focusing on the quasars and the few spectra that could not be
reliably classified. A comparison of the $K\le17.0$ quasar sample with
the $i$-band limited SDSS quasar catalogue is given in Section 5,
along with a discussion of the new quasars that possess unusual SEDs.
A discussion of the significance of the results and a comparison with
previous predictions appears in Section 6, with a summary following in
Section 7. Appendix \ref{appa} contains a description of the
non-quasars identified in the sample. Concordance cosmology with
$H_{0} = 70$ km s$^{-1}$ Mpc$^{-1}$, $\Omega_{m} = 0.3$ and
$\Omega_{\Lambda} = 0.7$ is assumed throughout.  Magnitudes on the
Vega system are used throughout the paper, with the SDSS AB-magnitudes
converted to the Vega system using the relations: $u=u_{\rm AB}-0.93$,
$g=g_{\rm AB}+0.10$, $r=r_{\rm AB}-0.15$, $i=i_{\rm AB}-0.37$, and
$z=z_{\rm AB}-0.53$, as described in Hewett et al. (2006). Unless
otherwise specified, the PSF magnitudes are used for the SDSS $ugriz$
bands, and the aperture corrected \texttt{aperMag3} magnitudes are
used for the WFCAM $YJHK$ bands.


\section{Sample Definition}


\subsection{UKIDSS}\label{UKIDSS}

The UKIRT Infrared Deep Sky Survey (UKIDSS, www.ukidss.org, Lawrence
et al.~2007) is a set of five separate surveys with complementary
combinations of areal coverage and depth.  Observations are carried
out on the UK Infrared Telescope (UKIRT) with the Wide Field Camera
(WFCAM) (Casali et al. 2007). The Large Area Survey (LAS) is the
largest of the surveys, and will cover 4000~deg$^2$ in $YJHK$ to a
5$\sigma$ limit of $K=18.2$, i.e. $\simeq 3$ magnitudes fainter than
2MASS. There have been three data releases to date, with the Early
Data Release (EDR, 2006 Feb, Dye et al. 2006), Data Release 1 (DR1,
2006 Jul, Warren et al.  2007a) and Data Release 2 (DR2, 2007 Mar,
Warren et al. 2007b), containing 27.3~deg$^2$, 189.6~deg$^2$, and
282~deg$^2$, respectively, of coverage in the full set of $YJHK$
bandpasses for the LAS.  Data is accessed via the online WFCAM Science
Archive (WSA\footnote{http://surveys.roe.ac.uk/wsa/pre/index.html},
Hambly et al. 2007). The LAS lies within the SDSS footprint and
each detected object will potentially possess $ugrizYJHK$ photometry
plus extensive morphological information. With such a wealth of
information available, a relatively sophisticated approach to the
definition of object samples for further study is possible.


\subsection{Initial Sample Definition}

Our study employs the UKIDSS EDR database and SDSS DR4 to define a
flux-limited object sample with $K\le17.0$. A list of objects
possessing measured $K$-band magnitudes was extracted from the UKIDSS
EDR and matched to the DR4 database using a matching radius of
$2.0\arcsec$, resulting in two sub-samples: one consisting of objects
detected in both UKIDSS and SDSS, and one with objects detected only
in UKIDSS. The matching objects with both sets of data comprise the
main sample to be used for further analysis, but both the matched and
UKIDSS-only object lists were retained.

The astrometric consistency between the SDSS and UKIDSS is better than
$0.5\arcsec$, and virtually no real matches are missed with a
$2\arcsec$ match radius. At matching radii $> 1\arcsec$, false
matches dominate, but an empirical determination of the frequency of
false positives shows that less than one percent of the matches are
coincidental.


\subsection{Quality Cuts}\label{cuts}

A magnitude limit of $K\le17.0$ was applied to both the matched and
unmatched object lists, the point at which the UKIDSS photometric
errors begin to increase significantly. For $K\le 17.0$, the error in
the $K$-band magnitude is $\sigma_{K}\le0.1$. A measured magnitude in
the $J$-band was also required of each object, effectively imposing a
$J$-band magnitude limit of $J\le19.6$. Objects detected within 64
pixels of the WFCAM chip edges were also removed as there are a large
number of spurious sources in these regions. A particularly effective
method of removing false UKIDSS detections is to cut out the regions
immediately surrounding bright stars and nearby galaxies. To obtain a
list of such objects, the 2MASS Point Source and Extended Source
Catalogs were queried for the areas of interest. Circular regions of
radius 1\arcmin\ around stars brighter than $K=9.5$ were cut out, as
were regions around galaxies of $K<12.0$ with their size determined by
the 2MASS-catalogued total radius. The Noise morphological
classification provided by the UKIDSS pipeline also eliminates a small
number of unmatched objects, but was not applied to the matched list.

The local sky variance values, provided in the UKIDSS database, for
each detection is an indicator of photometric quality. A histogram of
the measured variances for a given region displays a dominant, near
Gaussian, component with an extended tail toward large values. Objects
with sky variance values located in the tail lie in regions affected
by poor sky subtraction/determination, confirmed by visual inspection
of the images. Objects in the tail are excluded from the $K$-band
flux-limited sample because their photometry is adversely affected by
the sky-background uncertainties, which result in a noticeable
broadening of the stellar locus.  There remain a few objects whose sky
variance values are not extreme but whose WFCAM images indicate poor
sky subtraction, such as objects located within the very extended
halos of bright stars. There was no obvious method for removing these
objects without excluding significant areas of sky, so no further cuts
were imposed.

After all of these quality cuts, the remaining objects in the matched
list have good quality UKIDSS and SDSS photometry covering
$ugrizYJHK$, along with morphology classification, and in the case of
a few objects, SDSS spectra. Table \ref{tab:cutnum} lists the number
of matched and unmatched objects that remain in the sample following
the series of cuts for a typical 1.5~deg$^2$ area of sky.

The UKIDSS-only detections arise from several sources. First, there
are objects where the form of the SED is such that they are detected
in the $K$-band but not in $ugriz$. Reddened quasars, L or T dwarf
stars and elliptical galaxies above z $>1$ are examples of such
populations. Alternatively, the UKIDSS-only detection could be
spurious, as occurs frequently in the halos of bright stars or the
outer regions of large, nearby galaxies. Another common source of
UKIDSS-only detections arises from cross-talk between the detector
channels. For bright stars, cross-talk produces spurious images
located at integer multiples of 128 pixels from the star in all bands
(see Dye et al. 2006 for further explanation and examples). Asteroids
also contribute significantly to the UKIDSS-only population, but are
easily identified by shifts in position between the $K$ and $H$-band
images, which are always taken sequentially. For a small percentage of
sources, the SDSS photometric catalogue (PhotoObj) is incomplete, with
an object clearly visible in the SDSS imaging at the location of the
UKIDSS detection, but no photometry entry exists in the SDSS
catalogue.  Finally, pairs of objects with very small separations are
occasionally segmented differently by the UKIDSS and SDSS photometric
pipelines, producing mismatching entries in either the SDSS or UKIDSS
photometric catalogues. Objects from these five sources of spurious
images dominate the UKIDSS-only population, comprising $\sim$ 98 per
cent of the total.  Our primary goal of constraining the number of
heavily reddened quasars relies on identifying real UKIDSS-only
detections. In order to achieve the goal, both the UKIDSS and SDSS
images at the location of the UKIDSS-only detections were extracted
from the databases and inspected visually. The exercise was time
consuming but resulted in the elimination of all but the few objects
for which the UKIDSS detection is unambiguous and the SDSS image shows
no hint of a detection.

\begin{table}
\caption{\label{tab:cutnum}Number of objects that remain after each 
  cut for a typical 1.5~deg$^2$ area of sky.
}
\begin{tabular}{lcc} \hline
Cut & Matches & Unpaired \\ \hline
Initial Sample & 13404 & 1100 \\
$11.5<K<17.0$ & 5868 & 218 \\
Remove bright stars/galaxies & 5763 & 99 \\
Large sky variance & 5554 & 58 \\
Require $J$ and $K$ & 5536 & 58 \\ 
Right of selection line & 2335 & 58 \\
Centrally concentrated & 1378 & 58 \\
MergedClassStat$<$35 & 1214 & 58 \\
Visual inspection & 1214 & 1 \\ \hline
 & Stellar & Extended \\ \hline
Final Catalogue & 33 & 1182 \\ \hline
\end{tabular}\\
\end{table}


\subsection{Final Sample and Quasar Candidates}\label{qcandidates}

\begin{figure}
\resizebox{\hsize}{!}{\includegraphics{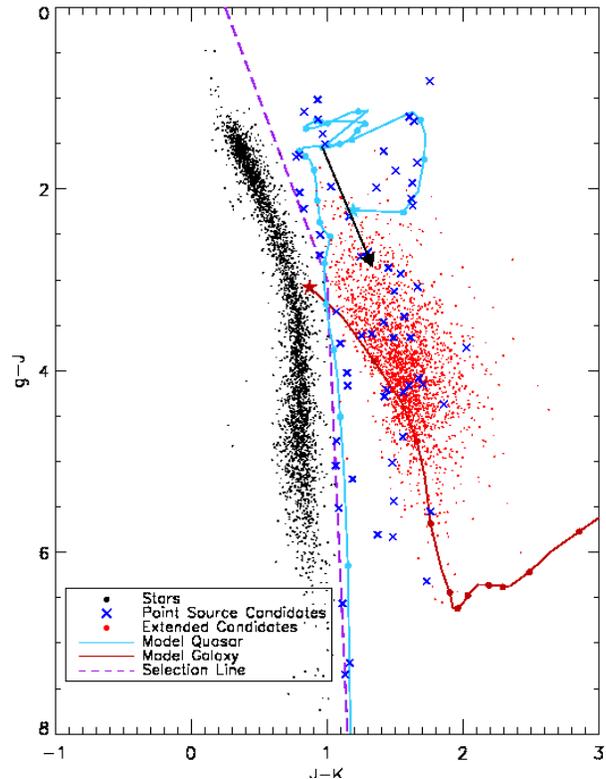}}
\caption{$gJK$ colour-colour plot for one 1.54~deg$^2$ field of combined
  UKIDSS LAS and SDSS data, showing all objects detected to $K\le17.0$.
  Black dots show the well-defined stellar locus, red dots show
  objects classified as extended, and blue crosses are objects
  classified as point sources. The blue track follows the $gJK$
  colours for a model quasar from $0<$ z $<4.8$, with the star
  indicating z $=0$ and dots at $\Delta$z $=0.2$ intervals. The red
  track follows the $gJK$ colours for a model elliptical galaxy for
  $0<$ z $<2.0$. The black arrow indicates the reddening vector for SMC
  reddening with E($B-V$) $=0.25$ at z $=1$. Every object lying to the
  right of the purple dashed line was considered as a candidate for
  observation, although not all were observed due to limited numbers
  of fibres.}
\label{fig:gJK1}
\end{figure}

The final $K$-band flux-limited sample of objects can be plotted on an
optical-NIR colour-colour diagram, exploiting the $K$-band excess of
quasars. Fig. \ref{fig:gJK1} shows a $g-J$ {\it vs} $J-K$ (hereinafter
$gJK$) diagram for one $\sim1.5$~deg$^2$ field suitable for follow-up
observation with the new dual-beam, multi-object AAOmega spectrograph
(Saunders et al. 2004) instrument on the Anglo-Australian Telescope
(AAT). Stars, which make up the majority of the sample, are confined
to a well-defined locus, as indicated by the black dots. The purple
dashed line is the selection boundary, described by:

\[
g-J=\left\{\begin{array}{ll}
    4(J-K)-0.6 & \mbox{for $J-K\le0.9$, $g-J\le3$} \\
    33.33(J-K)-27 & \mbox{for $J-K>0.9$, $g-J>3$} \\
\end{array}
\right.
\]

\noindent Objects to the left of the line are eliminated, with objects
to the right considered to be a quasar candidate. The
position of the selection line along the $J-K$ axis varies slightly
from field to field (the hinge point changes from $J-K=0.9$ to
$J-K=1.0$) to account for variation in photometric quality. The
stellar locus is broader in fields with poor photometry, forcing the
selection line rightward to prevent the quasar-sample from being
overwhelmed by stars scattered across the selection boundary.  

The blue track follows the $gJK$ colours for a parametric quasar
model, the components of which have been optimised to reproduce the
$ugrizYJHK$ colours of the matched SDSS--UKIDSS spectroscopic quasar
sample over $0.0<$~z~$<4.8$. The model is similar to that described in
Maddox \& Hewett (2006, Section 2.2), with some alterations. The
power-law component is now described by $\alpha=-0.54$ for
$\lambda<2750$\,\AA, and $\alpha=0.041$ for $\lambda>2750$\,\AA, where
$F(\nu) \propto \nu^\alpha$. The upturn at NIR wavelengths is provided
by a blackbody with temperature 1775~K, similar to the composite
spectrum of Glikman, Helfand \& White (2006).  For the model to match
the median colours of matched SDSS-UKIDSS quasars at low redshifts
(Chiu et al. 2007), flux from an Sb-type host galaxy is included. The
contribution from the host galaxy decreases rapidly at increasing
redshift. The red track follows the $gJK$ colours for a model
elliptical galaxy as described in Mannucci et al.  (2001) for $0<$ z
$<2.0$. The `star' symbols on both the quasar and galaxy model tracks
indicate redshift zero.

Use of the $g$-band in the colour selection restricts the redshifts
accessible by this technique to z~$<4$, as there is little flux
falling within the $g$-band for higher redshift objects. This effect
is evinced in Fig. \ref{fig:gJK1} by the model quasar track rapidly
becoming very red in $g-J$. The two-colour selection technique can be
easily extended to higher redshifts by using redder optical passbands,
for which the diagrams are similar.

The objects in Fig. \ref{fig:gJK1} are segregated by their UKIDSS and
SDSS morphological classifications into unresolved and resolved
sources. The SDSS morphology is determined by the difference between
the PSF and the CModel magnitudes\footnote{See
  http://www.sdss.org/dr5/algorithms/ for definitions of PSF and
  CModel magnitudes}, whereas morphology classification for UKIDSS
data is derived from the flux curve-of-growth of each object.  Blue
crosses are candidates with point source morphology as determined by
both UKIDSS and SDSS, which constitute only a small fraction of the
initial sample, and are assigned the highest observing priority value
of 9. Quasars included in the SDSS DR3 Quasar Catalog (Schneider et
al. 2005) were also specifically included as targets for observation.
Six objects with secure UKIDSS identifications and no SDSS
counterpart were also assigned high observing priority.

In Fig. \ref{fig:gJK1}, red dots indicate candidates with extended
morphology. As quasars reside in host galaxies, which may contribute
significantly to the total flux from the object at longer wavelengths
(Maddox \& Hewett 2006), extended objects should not be simply
excluded. As the goal of this study is to create a \emph{quasar}
sample to $K\le17.0$, it is important to differentiate between the
nuclear and host galaxy flux. The WFCAM pipeline provides magnitude
measurements derived from the flux falling within apertures of
different radii. \texttt{aperMag3} is recommended as the aperture
providing the most accurate and stable measure of magnitude for a
variety of objects, and includes flux to a $1.0\arcsec$
radius\footnote{Recall that \texttt{aperMag3} is used to define the
  base $K\le17.0$ flux-limited sample}. \texttt{aperMag1} is based on
flux falling within the innermost $0.5\arcsec$ radius and provides a
bright limit to the flux from any unresolved nuclear source present in
resolved objects. A more sophisticated estimate of the nuclear
$K$-band flux present in each resolved object is provided in Section
\ref{exflux}. However, for the purposes of defining the sample for
spectroscopic observation, only resolved objects with
$K_{\texttt{aperMag1}}\le17.0$ were included (listed as `Centrally
concentrated' in Table \ref{tab:cutnum}). Very extended, nearby
galaxies, with UKIDSS MergedClassStat (a measure of stellarness) $>35$
were also excluded.

Even after applying the restriction on nuclear flux, there are still
far more extended objects than can be observed with the $\simeq300$
fibres available for each 1.5~deg$^2$ field.  Instead of employing a
straightforward 1 in N sampling of the extended objects, their
distribution in $gJK$ colour space is used.  Approximately 120
extended objects in the most sparsely populated regions of colour
space are assigned observing priority values of 8.  The next 250
objects with increasing density are retained and assigned priority
values of 6, and the next approximately 50 objects are retained and
assigned values of 2. This provides a list of $\sim$400 extended
objects, in addition to the $<100$ point sources, from which the
2-degree Field (2dF, Lewis et al.  2002) fibre configuration tool can
choose. An observing strategy that maximised the total area of sky covered
was chosen over multiple telescope pointings of
the same field because all of the high-probability quasar candidates,
i.e. the stellar objects, in a given field could be
observed in a single pointing. The acquisition of additional spectra of
extended objects would result in a more complete census of low-luminosity
AGN but this was not a key objective of our investigation..

\section{New Spectroscopic Data}


\subsection{Observations}

Spectroscopic observations were performed on 2006 April 30, May 1--3,
at the Anglo-Australian Telescope. A summary of the fields observed is
provided in Table~\ref{tab:obs}. The last three fields listed in
Table~\ref{tab:obs} were observed in poor conditions. Spectra were
obtained using the 2dF fibre positioning robot and the AAOmega
spectrograph. The 580V and 385R VPH gratings were used, providing
spectral coverage 3700-5800\,\AA \ and 5600-8800\,\AA \ in each arm,
with the two segments spliced together between 5600 and 5800\,\AA \ 
giving continuous coverage at resolution 1300 from 3700-8800\,\AA. Of
the 400 fibres, eight are allocated for guide stars, 35-40 are used as
sky fibres, and approximately 50 fibres were damaged, leaving just
over 300 fibres available for science targets.  Eleven fields were
observed in total, resulting in 3154 usable spectra spanning
approximately 14.2~deg$^2$, making this study the largest $K$-band
selected quasar survey to date by a factor of more than twenty.
Accounting for the area lost due to the quality cuts described in
Section \ref{cuts}, and the damaged fibres outlined in Section
\ref{noids}, the effective area of the survey is
12.8~deg$^2$.

A subset of the fibres exhibit a fringing pattern, caused by a gap
between the front end of the fibres and their retractors. The fringing
is not stable, resulting in unusable spectra in some cases. In the
less severe cases, object identification is still possible, but
measurements of linewidths or continuum values are not. In order to
reduce the impact of these fibres on the experiment, the highest
priority targets (i.e.  point sources) were first assigned to fibres
known not to suffer from fringing. Lower priority objects were
assigned randomly to the remaining fibres, which include both fringing
and non-fringing fibres. Unfortunately, the available list of fibres
experiencing fringing was not complete and occasionally a high
priority target was assigned to a fringing fibre. The (small) number
of fibres affected by fringing is listed in the last row of
Table~\ref{tab:res}.

\begin{table}
\begin{center}
\caption{\label{tab:obs}AAOmega observation log.}
\begin{tabular}{ccccc} \hline
RA$^a$ & Dec.$^a$ & Seeing & Exposure time & Date \\ 
(J2000) & (J2000) & (\arcsec) & (s) & (yyyy-mm-dd) \\ \hline
185.10 & 0.0 & 1.6 & 3600 & 2006-05-02 \\
189.90 & 0.0 & 1.6 & 3600 & 2006-05-03 \\
194.00 & 0.0 & 1.6 & 4200 & 2006-05-02 \\
195.70 & 0.0 & 1.5 & 3600 & 2006-05-03 \\
197.40 & 0.0 & 1.3 & 4100 & 2006-05-03 \\
199.10 & 0.0 & 1.3 & 6600 & 2006-05-02 \\
204.20 & 0.0 & 1.3 & 4800 & 2006-05-02 \\
212.70 & 0.0 & 1.8 & 3900 & 2006-05-02 \\
214.40 & 0.0 & 2.0 & 4500 & 2006-05-02 \\
236.00 & 5.7 & 2.3 & 3600 & 2006-05-03 \\
237.75 & 5.7 & 2.5 & 4800 & 2006-05-02 \\ \hline
\end{tabular}\\
\end{center}
$^{a}$ {\footnotesize Centre of each observed field}\\
\end{table}


\subsection{Data Reduction}

All spectra were reduced with the custom reduction package 2dF Data
Reduction (2dFDR, Bailey et al. 2003), upgraded to handle the separate
blue and red AAOmega spectrograph arms and splice the reduced spectra
together. The classification routine \texttt{RUNZ} (Colless et al.
2001) was used in manual mode to identify the quasars, star-forming
and absorption line galaxies, and stars. The classifications and
redshifts for each object were determined by eye.


\subsubsection{Spectra Quality}\label{response}

The new AAOmega spectra of confirmed quasars from the SDSS DR3 Quasar
Catalog allow an assessment of the effectiveness of our quasar
identification and provide a direct empirical determination of the
AAOmega wavelength dependent sensitivity. There does not appear to be
any significant variation in wavelength response dependent on fibre
location within the field, although some fields have a more consistent
response than others. The last two fields listed in Table
\ref{tab:obs} had no SDSS spectral coverage at the time of
observation, but the area has since been observed and included in the
newly released SDSS DR5 Quasar Catalog (Schneider et al. 2007).
Confirmed KX-selected quasars were matched to the SDSS quasars with a
matching radius of $1.0\arcsec$.

As these observations were some of the first performed with AAOmega,
the 2dFDR splicing routine employed to join the red and blue arms was
still under development. The splicing routine was performing poorly in
some cases, particularly for objects with low signal, or when an
emission line was present in the joining region.  The red and blue
arms for each spectrum are reduced independently, and are thus
available for additional manipulation in order to improve poor joins.

The SDSS spectra and the new observations can be used to determine the
AAOmega spectral response and correct poor splicing. 89 such spectra
exist, with 78 below z $<2.1$. High redshift objects were excluded as
the Ly$\alpha$ forest means there is very little flux in the blue arm.
After median filtering each spectrum with a 15 pixel window to remove
noise spikes but retain spectral features, the SDSS and blue arm
AAOmega spectra were normalised between 4000-5000\,\AA \ then divided,
and the SDSS and red arm AAOmega spectra were normalised between
6000-7000\,\AA \ then divided, to provide the red and blue response
curves for each AAOmega spectrum.  The median red and blue response
curves for each field was then computed.

\begin{figure}
\resizebox{\hsize}{!}{\includegraphics{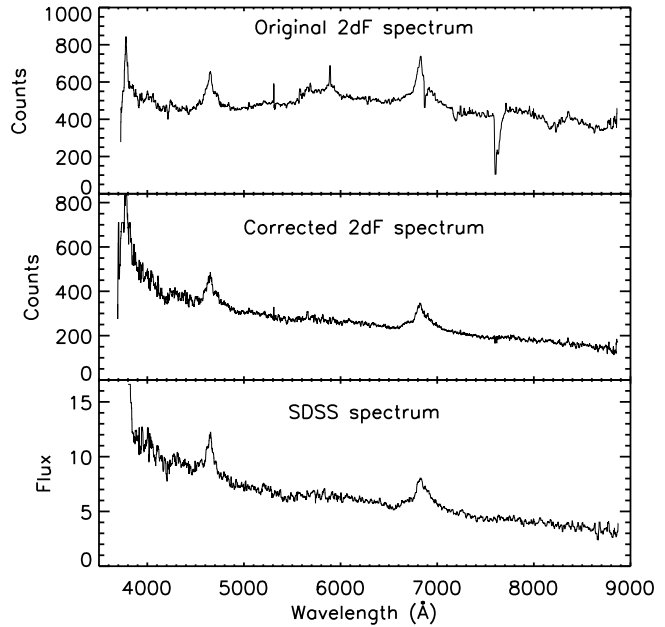}}
\caption{Illustration of the effectiveness of the wavelength dependent
  response correction. (Top) The raw AAOmega spectrum of a quasar at z
  $=1.44$, as produced by the 2dFDR pipeline.  (Middle) Corrected
  spectrum, derived using the procedures described in the text. The
  continuity of the corrected spectrum in the region where the red and
  blue arms are spliced together (5600-5800\AA) improves
  significantly, and the shape of the corrected spectrum more closely
  matches the fluxed SDSS spectrum of the quasar (Bottom). Note that
  the telluric A and B absorption bands visible in the original
  spectrum have been removed by the response-correction process. The
  AAOmega spectra have been median filtered using a 15-pixel window,
  as described in the text.}
\label{fig:correct}
\end{figure}

After applying the median response curves to the separate red and blue
spectral arm data, the arms can be spliced together. As a test, the
median response curves were applied to the original individual AAOmega
spectra that have corresponding SDSS spectra. The procedure produces a
significant improvement in both overall shape and in continuity across
the splice region; an example is shown in Fig. \ref{fig:correct}. The
median response curves were then applied to all spectra containing
emission lines.

The response correction was applied after the spectroscopic
classifications were performed, to provide an approximate relative
flux calibration for each spectrum. Had the response correction been
applied before the spectral classification, the identification success
rate would not have increased significantly. The insensitivity of the
classification success rate to the application of the response
correction is due to the high fraction of spectra identified and the
properties of the unidentified spectra, which nearly always suffer
from very poor signal-to-noise ratio.

\subsection{Classification}\label{classsec}

Although many spectra allowed high confidence identifications to be
made, a number of spectra were not easily identifiable due to low
signal-to-noise ratio (SNR) or a lack of obvious spectral features.
For the eight fields with the best observing conditions,
$<$10~per~cent of the spectra remained unclassified. When the three
poorest quality fields are included, the number of unclassified
spectra increases to 14~per~cent. Of the 3154 new AAOmega spectra,
2728 resulted in classifications. The other 426 objects for which
spectra exist but didn't result in a classification are discussed in
Section~\ref{noids}.


\subsubsection{Spectroscopically Unclassified Objects}\label{noids}

\begin{figure}
\resizebox{\hsize}{!}{\includegraphics{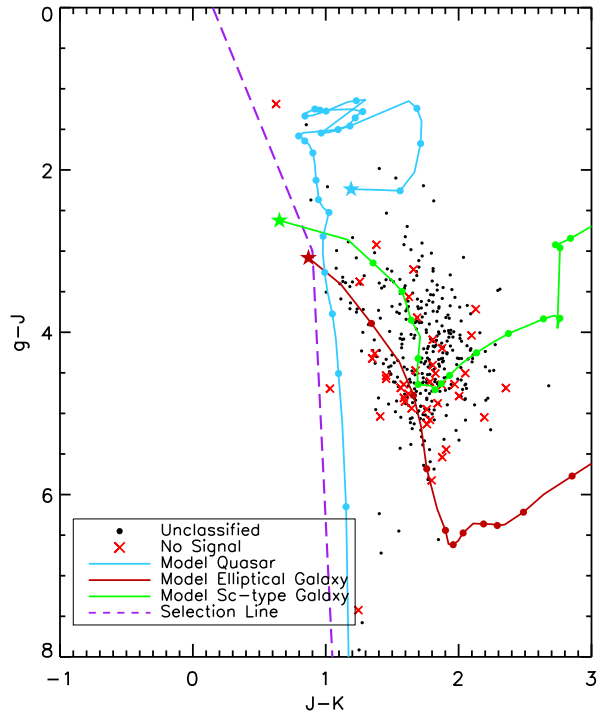}}
\caption{Location of spectroscopically unclassified objects on the
  $gJK$ diagram. Black dots indicate objects whose spectra have flux 
  but do not contain any identifiable features, and red crosses mark
  objects whose spectra do not contain any significant flux. The
  purple selection line, blue model quasar and red elliptical galaxy
  tracks are as in Fig.  \ref{fig:gJK1}, with the green track
  following the colours for a model Sc-type galaxy.}
\label{fig:noid}
\end{figure}

Spectra that did not produce an identification fall into three
classes. First, there are objects with anomalously low SNR spectra,
given their SDSS magnitudes, due to some instrument-related problem
unrelated to the intrinsic nature of the objects. These objects are
distributed essentially randomly in $gJK$ colour space, as shown by
the red crosses in Fig.  \ref{fig:noid}. Second, the fibre may be
suffering from severe fringing, making an identification impossible.
These spectra are referred to as `Damaged' in Table \ref{tab:res}.
Fibres that exhibit the fringing pattern were preferentially assigned
to extended objects as they were considered to have lower observing
priority than point sources. In a number of cases, an identification
was still possible, due to strong emission lines or only mild
fringing. These spectra are marked as having an identification, but
they are not included in the `Best' category in Table \ref{tab:res}.
The 43 low SNR spectra and 56 fringing spectra decrease the effective
area of the survey by 0.4 deg$^2$, leaving 12.8~deg$^2$ total
effective area, and are not included in subsequent analysis of
unclassified spectra.

The remaining unclassified objects possess spectra of reasonable SNR
but do not show any identifiable features. The location of these
objects on the $gJK$ diagram is shown by the black dots in Fig.
\ref{fig:noid}. In addition to quasar and elliptical galaxy
model tracks, the model track for an Sc-type galaxy from Mannucci et
al. (2001) is overlaid on the $gJK$ diagram. The unidentified objects
are primarily located between the elliptical and Sc-type galaxy
tracks. The redder objects cluster around the elliptical galaxy track
at z $\simeq 0.4$. It is difficult to identify with confidence an
absorption line galaxy at this redshift, as the 4000\AA-break falls in the
region of the spectrum where the AAOmega red and blue spectral arms
are spliced together. The cluster of objects slightly bluer in $g-J$
occupy the region of colour space dominated by star-forming galaxies
at a range of redshifts. Emission lines of only moderate strength can
be difficult to identify given the modest SNR of the spectra of some
of the fainter objects.  The few objects located at $g-J\simeq 6.5$
are probably faint M-type stars, whose red SEDs are difficult to
identify unambiguously given the poor sky subtraction at $>$7500\,\AA
\ in the AAOmega spectra.  There are very few unclassified objects
located near the quasar track for $z<3$ on the $gJK$ diagram,
indicating that the identification success rate for unobscured quasars
is very high.

The distribution of unidentified spectra in the $gJK$ diagram accords
well with expectations. For $g-J<3$, where the majority of objects are
quasars or AGN with prominent emission lines, 823 objects were
observed, 797 have classifications, 9 were affected by
damaged/fringing fibres, leaving only 17 objects with reasonable
signal-to-noise ratio and no classifications.  For $g-J>3$, where the
majority of galaxies lie, including many with no readily identifiable
emission lines, 2331 objects were observed, 1931 have classifications,
90 were affected by damaged/finging fibres, leaving 310 with no
classifications. The difference in classification success, 97~per~cent
($g-J<3$) compared to 83~per~cent ($g-J>3$) is primarily attributable
to different object populations present in each colour interval. The
photometric classification statistics presented in
Section~\ref{photoz1} confirm such a conclusion.

The location of an object on the $gJK$ diagram provides a clue to
the classification of the object. However, there is an overlap between
the location of the many galaxies and the few highly reddened (with
E($B-V$) $>0.5$) quasars, at $g-J\simeq4$ and $J-K\simeq1.8$. As a
consequence, we employ photometric redshift techniques, using the full
$ugrizYJHK$ photometry, and the morphological information available for
each object, to provide high-confidence identifications for the objects
without spectroscopic classifications. 


\subsubsection{Photometric Redshifts}\label{photoz1}

Although nearly all (93\,per cent) of the stellar candidates possess
spectra, thousands of extended objects with $K\le17.0$ could not be
observed due to the finite number of spectroscopic fibres available.
An additional 516 spectra obtained as part of the SDSS spectroscopic
programme, consisting mostly of galaxies, are added to the 2728 of
3154 AAOmega spectra which resulted in classifications to increase the
number of spectroscopically classified objects to 3244. For objects
with no spectroscopic information, the $ugrizYJHK$ photometry, plus
the morphological information, available for each object is used
instead to provide classifications. At this stage we are interested in
establishing whether the SEDs of the candidates without spectroscopic
classifications are consistent with those of ordinary stars or
galaxies. We return to the more ambitious goal of establishing that
each of the candidates is not consistent with harbouring a reddened
quasar with $K\le17.0$ in Section \ref{exflux}.

The SDSS has determined photometric redshifts and type classifications
for every object contained within the DR5 catalogue by fitting
template galaxy SEDs to the observed objects' $ugriz$ photometry (for
detailed information regarding the Photoz algorithm, see Csabai et al.
2003). The SDSS Photoz is optimised for use with galaxies, therefore
quasars and stars generally result in unreliable fits, as determined
by their large $\chi^2$ and redshift error values. The results for
each object identified as a potential observation candidate (objects
lying to the right of the selection line in a $gJK$ diagram) were
extracted from the SDSS Photoz table. An object with a good Photoz
template fit, as determined by a value of $\chi^2<20$ or a quality
flag of either 4 or 5, can be considered to have an SED consistent
with that of a normal galaxy, and not a quasar.

For the small percentage of objects which result in a poor photometric
redshift fit and have no spectroscopic classification, further
investigation into their possible identification is required. In a
manner similar to the SDSS Photoz algorithm which uses only the
optical magnitudes and galaxy SEDs, model SEDs extended to the NIR can
be fit to the full suite of $ugrizYJHK$ photometry, with the
expectation that the additional passbands will provide extra
discriminatory power. In addition to the $YJHK$ photometry, model
templates of an unreddened quasar, five reddened quasars with
$0.1<$~E($B-V$)~$<1.0$, several galaxy types and galactic stars are
included in the fitting, with the aim of distinguishing between the
various types of objects that may be present in the whole population,
not just identifying the galaxies. For clarity, the photometric
redshift results provided by the SDSS will be referred to as SDSS
Photoz, and the photometric redshift results using the additional NIR
photometry and quasar templates will be referred to as the
NIR-extended Photoz. As a consistency check, both photometric redshift
schemes were performed on the set of objects with spectroscopic
identifications and redshifts to ensure reliable results are computed.
The results from both schemes, when applied to the entire population
of extended candidates, are also highly consistent. Plots displaying
the quality of the photometric redshifts and spectral types are
provided in Appendix~\ref{appb}.

Combining the SDSS Photoz and the NIR-extended Photoz results with the
new and existing spectroscopic data greatly increases the number of
objects for which a confident classification exists, as for the
objects with spectra, or whose colours are consistent with those of
ordinary galaxies. Table~\ref{tab:photoz} lists the order in which
objects were identified and subsequently removed from the initial
population of 11\,895 objects located to the right of the selection
line in the $gJK$ diagram for the 13.8~deg$^2$ area. Objects with
spectroscopic identifications are considered to be the most secure,
and are thus removed first. Next, objects that are classified as
morphologically stellar and are classified by the NIR-extended Photoz
as stars are removed. Objects with high confidence SDSS Photoz results
are then removed, which includes the bulk of the remaining population.
The NIR-extended Photoz removes a further five objects, leaving only
eleven objects remaining without classification of any kind.  The SED
of one of the five objects removed is consistent with that of an
unreddened quasar, and is flagged as 'TargetQSOFaint' by the SDSS, but
was not observed spectroscopically.  Inspection of the eleven
remaining images reveals six of these objects to be apparent close
pairs of objects for which there is only one entry in the SDSS
photometric database, one object is a saturated star, and the
remaining four objects are the UKIDSS-only detections, discussed
further in Section \ref{nosdss}.

\begin{table}
\centering
\caption{\label{tab:photoz}Number of unidentified objects left after
  each identification routine is applied}
\begin{tabular}{lcc} \hline
Identification & Removed & Remaining \\ \hline
Initial Sample Size & 0 & 11895 \\
Spectral ID & 3244 & 8651 \\
Stellar & 10 & 8641 \\
SDSS Photoz & 8625 & 16 \\
NIR-extended Photoz & 5 & 11 \\ \hline
\end{tabular}\\
\end{table}

Focusing on the subset of 327 objects observed with AAOmega that did
not result in spectroscopic identifications outlined in Section
\ref{noids}, only five objects did not result in a good fit from the
SDSS Photoz or the NIR-extended Photoz and are a subset of the eleven
unclassified objects described above. The computed redshift
distributions for the 322 classified objects span $0<$~z~$<1.0$, and
peak at z~$\simeq0.4$, consistent with the conclusions drawn from the
$gJK$ plot. The type distributions, available from SDSS Photoz and the
NIR-extended Photoz, contain a majority of early type galaxies, with a
significant population of galaxies with a range of star-formation
activity. The SDSS Photoz results for these 322 objects are shown in
Fig.~\ref{fig:pz3} of Appendix~\ref{appb}.

Applying the NIR-extended photometric redshift scheme to the entire
population of candidates results in a significant number of objects
that are classified as ordinary galaxies with high confidence, but
whose colours are also consistent with the reddened quasar models.
The redshift distribution from the galaxy classifications is very
similar to the redshift distribution of the spectroscopically
confirmed galaxy population, whereas the redshift distribution from
the reddened quasar classifications shows a sawtooth pattern with
peaks at z = 0.3, 1.3 and 2.3. Based on these redshift distributions,
it was assumed that in each case, the `galaxy' classification was
correct. However, as described in Section \ref{exflux}, the
possibility of these objects harbouring reddened quasars bright enough
for inclusion in a $K\le17.0$ sample is investigated and ruled out.


\subsection{Excess Nuclear Flux in Extended Objects}\label{exflux}

As seen in Section \ref{photoz1}, all but eleven objects of the
initial 11\,895 candidates possess either spectroscopic or confident
photometric classifications. However, for 2249 of the objects, the
photometric information is not sufficient to exclude the possibility
that the object may be a reddened quasar. In this section we use a
more sophisticated approach, based on the radial light profile
information available in SDSS and UKIDSS for extended objects, to
eliminate objects as candidate reddened quasars.  Specifically, for
all but a handful of objects, we show that any unresolved nuclear
component present in the $K$-band is fainter than the $K=17.0$
flux-limit of our quasar sample.

We have already applied a morphological restriction to eliminate very
low surface brightness galaxies from the sample targeted for
spectroscopy. The selection (Section \ref{qcandidates}) required that
the $K$-band flux from the innermost $0.5\arcsec$ radius satisfied
$K\le17.0$. However, this measure includes both flux from the inner
regions of the host galaxy in addition to any nuclear source that may
be present. A more robust method of measuring flux from a nuclear
point source that accounts for flux from the host galaxy is possible
using the radial surface brightness profile information provided for
each object by the SDSS and the UKIDSS databases. For the procedure
described below, the SDSS $i$-band data is chosen because the $i$-band
is the reddest passband for which the SDSS imaging is of high SNR and
the galaxy radial profiles in the $i$ and $K$-bands are not expected
to be significantly different. Galaxy profile fits are not provided as
part of UKIDSS but we use the $K$-band flux in two different apertures
to constrain the nuclear flux.

For each object, the SDSS data processing pipeline fits both a
deVaucouleurs and an exponential radial surface brightness profile in
each photometric band, and provides the likelihood of the object being
a star or a galaxy, of either profile, based on which fit best matches
the data. Objects for which none of the profiles are good fits, such
as merging galaxies, have very small likelihoods for all three fits.
Using the SDSS $i$-band profiles, the candidate list of 11\,895
objects is split into four categories: stellar, deVaucouleurs,
exponential or no adequate fit. For each extended object with a good
profile fit, the SDSS model profiles are blurred, using the UKIDSS
$K$-band seeing, to predict the ratio of fluxes in two apertures of
radii $1.0\arcsec$ and $2.83\arcsec$, corresponding to the UKIDSS
\texttt{aperFlux3} and \texttt{aperFlux6}. In other words, based on
the SDSS-provided model profile fits and the $K$-band seeing, it is
straightforward to predict the aperture flux ratio of a galaxy with no
additional central point source.  Then, using the measured $K$-band
values of \texttt{aperFlux3} and \texttt{aperFlux6}, a central flux
excess with respect to the predicted value can be computed. The excess
flux, if present, may be converted to an excess magnitude (including
appropriate aperture and zero point corrections), and if this excess
magnitude is brighter than $K=17.0$, the galaxy hosts a nuclear
source bright enough to be included in a $K\le17.0$ survey.

The model predictions showed good agreement with the observed $K$-band
flux ratios for the galaxy population. However, as the number of
objects available is large, the reference flux ratio for galaxies
(without any additional nuclear component) was derived as follows. For
each profile type (deVaucouleurs or exponential), the median flux
ratio as a function of profile scale-length, either the half-light
radius or exponential scale length, was determined for all resolved
objects observed in similar seeing conditions in the $K$-band. Seeing
intervals of $<0.7\arcsec$, 0.7--0.8, 0.8--0.9, \ldots, 1.1--1.2,
$>1.2\arcsec$ were used. The behaviour of the aperture flux-ratios was
systematic and well-defined over the full range of profiles observed
in each seeing interval. The excess nuclear flux for each object was
then calculated from the observed flux excess over and above the
empirical median flux ratio for the object's profile scale-length.

Focusing on the 2249 objects whose SEDs are consistent with both
ordinary galaxies and reddened quasars, only two are computed to have
nuclear magnitudes brighter than $K\le17.0$. The first, ULAS
J125949.08+001344.9, is the smaller of a close pair of possibly
interacting galaxies, and thus the radial light profile is most likely
affected by the neighbour. Although the NIR-extended Photoz initially
classified this object as a reddened quasar at redshift z $=2.7$, the
SDSS Photoz classifies it with very high confidence as an ordinary
galaxy at z~$=0.14$. The SDSS spectrum of the companion galaxy shows
evidence of star-formation, and lies at z~$=0.13$, lending further
confidence to the SDSS Photoz result.

The second object, ULAS J131101.23+000310.8, is a resolved galaxy. The
NIR-extended Photoz classified this object as a highly reddened quasar
at z~$=0.1$, but the SED is also consistent with an Sbc-type galaxy at
z~$=0.15$. The SDSS Photoz results agree with the classification of a
star-forming galaxy. The existing SDSS spectrum shows narrow emission
lines with type 2 ratios at z $=0.096$, indicating the possible
presence of an obscured active nucleus. However, its low redshift
ensures that the object would not satisfy the absolute magnitude
criterion of M$_i<-22.4$ required for inclusion in the quasar sample.

Objects classified as stellar are not included in this analysis, as
virtually every stellar candidate was observed spectroscopically. The
galaxy profile-based procedure is not expected to work for low
redshift, extended objects with a quasar visible in the $i$-band, as
their SDSS surface brightness profiles should not be well fit by
either a deVaucouleurs or an exponential model. However, the targets
of our investigation are galaxies hosting optically obscured quasars,
where the obscured quasar is not evident in the SDSS $i$-band. Thus,
the SDSS galaxy surface brightness profiles should be largely unaffected by
the hidden quasar and provide a good estimate of the expected $K$-band
\texttt{aperFlux3}$/$\texttt{aperFlux6} ratio.  Then, due to the much
reduced extinction in the $K$-band, a central flux excess should be
apparent if a bright nuclear source is present.

In summary, although not every object located to the right of the
selection line on a $gJK$ plot was observed spectroscopically, we have
used all of the available photometric and morphological information
from both the SDSS and UKIDSS databases to assign a high confidence
classification to each object. The cases for which the photometry is
consistent with that of a reddened quasar, the radial light profile
information is used to exclude the possibility of the objects
harbouring nuclear point sources bright enough to be included in our
$K$-band $K\le 17.0$ sample. Only eleven objects remain without such
classifications, and in each case the reason is readily understood.


\section{$K$-band Selected Sample}


\subsection{Sample Composition}\label{sample}

\begin{table}
\centering
\caption{\label{tab:res}Spectroscopic classification of objects. The
  second column includes data from all 11 observed fields with
  confident and very confident identifications, while the third column
  only includes data from the eight fields with the best observing
  conditions and initial photometry, and only very confident identifications.}
\begin{tabular}{lcc} \hline
Category & All & Best \\ \hline
Broad-line Quasars/AGN & 196/15 & 137/6 \\ 
Narrow-line Type 2 & 96 & 74 \\ 
Star-forming Galaxies & 944 & 676 \\
Absorption line Galaxies & 1206 & 718 \\
Stars & 270 & 172 \\
Spectroscopically Unclassified & 327 & 159 \\ 
Damaged & 99 & 45 \\ \hline
\end{tabular}\\
\end{table}

\input{table5.tex}

Each of the 3154 AAOmega spectra were classified as either broad-line
quasars/AGN, narrow-line type 2 Seyfert galaxies, star-forming
galaxies with obvious emission lines, absorption line galaxies, stars,
or remained spectroscopically unclassified. The number of objects in
each category is listed in Table \ref{tab:res}. The first column lists
the type classification, the second column includes every
object of each type, while the third column only includes objects
with very high confidence identification and whose spectra are of high
enough quality for further study, i.e.  excluding fields observed
under poor conditions, for example. One BL Lac object was identified
and excluded from the quasar sample. Details for the broad-line
objects follow in Section \ref{quasars}, whereas the unclassified
objects were examined in Section \ref{photoz1}. Information for the
other object classes may be found in Appendix \ref{appa}.

Table~\ref{tab:allobs} shows a subset of the object catalogue. The
complete catalogue appears \underline{in the online version} of this
paper, along with a description of each column. Descriptions of only
the columns appearing in the table subset are given here. Column~1 is
the IAU name of each object. Column~2 contains the UKIDSS morphology
class, with 1~=~extended, 0~=~noise, -1~=~stellar, -2~=~marginally
stellar, -3~=~marginally extended, -9~=~saturated. Columns~3--4
contain the SDSS PSF $g$-band magnitude and error,
while Columns~5--8 contain the UKIDSS \texttt{aperMag3} $J$ and
$K$-band magnitudes and errors.

Column~9 lists the E($B-V$) values provided in the WFCAM Science
Archive, derived from the galactic dust extinction value measured from
the Schlegel, Finkbeiner \& Davis (1998) maps. The identifying type
name in Column~10 is derived from the spectroscopic identifications,
with `QSO' for broad-line quasars and AGN, `Abs' representing all
absorption line galaxies, `Em' for star-forming galaxies, and `Type2'
for emission-line galaxies showing type 2 emission line ratios. `NoID'
denotes objects with flux in their spectra but no identifiable
features, `Fringing' is for objects in which the fringing within the
fibre prohibited an identification, and `NoFlux' identifies spectra
with particularly low SNR. The stars are divided into three broad
classes of `Astar', `Kstar', and `Mstar', and `BLLac' denotes the one
object whose spectrum was entirely featureless. Column~11 provides the
spectroscopic redshifts, which is set to be 0.000 for stars and
\mbox{-99.999} for objects with no identification. Column~12 contains an
approximate measure of the SNR of each spectrum, computed from the
mean and standard deviation of the spectra between 6000--7000\AA.

Of the six objects that were observed specifically because
they had detections in UKIDSS but not in any of the SDSS bands, two
were classified as absorption line galaxies, one has some flux in its
spectrum but has no identifiable features, and three do not have
sufficient flux in their spectra to make identifications. These will
be discussed further in Section \ref{nosdss}.


\subsection{Quasars and AGN}\label{quasars}

Quasars and AGN are separated from other objects by requiring the
presence of one emission line of full width at half maximum (FWHM) of
at least $1500$ km s$^{-1}$, with the distinction between quasars and
AGN based purely on an absolute magnitude cut of M$_i < -22.4$. In
practice, it does not make a significant difference if the broad line
cutoff is reduced to $1000$ km s$^{-1}$, as only a handful of objects
are added, some of which exhibit type 2 high ionisation emission line
ratios. One object is added to the broad line sample by hand, due to
the presence of very broad wings at the base of strong narrow lines.
Line characteristics for the quasars and all of the emission line
objects are measured using the IDL program line\_eqwidth.pro from the
FUSE IDL Tool
package\footnote{http://fuse.pha.jhu.edu/analysis/fuse\_idl\_tools.html}.
The measurements are approximations only, as values such as FWHM are
estimated assuming the emission line follows a Gaussian profile.

The absolute magnitude limit is taken directly from the SDSS quasar
selection algorithm (Richards et al. 2002), converted to the Vega
system. After correcting the apparent magnitude measurements for
Galactic extinction using the maps of Schlegel et al. (1998), absolute
magnitudes for the broad-line objects are calculated employing
K-corrections computed from the quasar+galaxy composite model
described in Section \ref{qcandidates}. As the model quasar is
representative of a typical, unreddened quasar, the calculated
absolute magnitudes are upper limits (i.e. lower limits on the
luminosities).  Using models that include dust reddening or more host
galaxy flux will serve to increase the K-correction, thus decreasing
(i.e.  brightening) the computed absolute magnitudes.  15 of the 211
broad-line objects fail to meet the M$_i < -22.4$ restriction when
using the unreddened quasar model K-correction, and are confined to z
$<0.6$. These 15 objects will hereinafter be referred to as AGN, not
quasars.

\begin{figure}
\resizebox{\hsize}{!}{\includegraphics{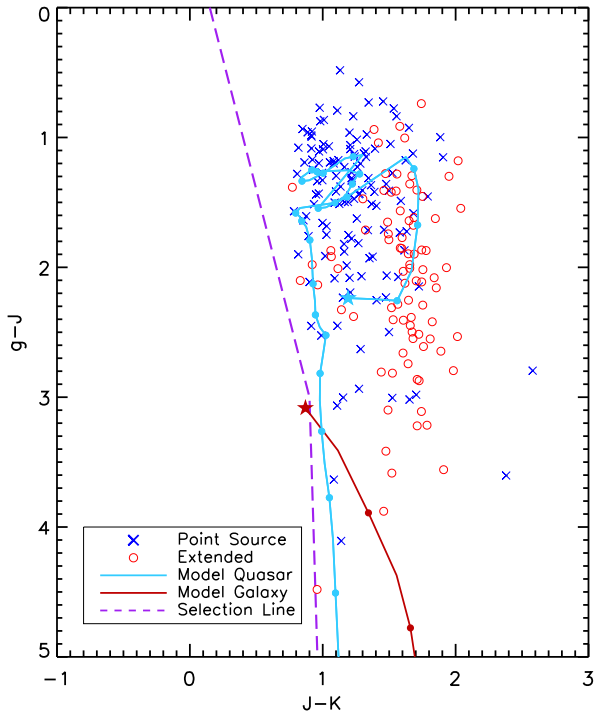}}
\caption{$gJK$ colour-colour diagram with confirmed broad-lined
  quasars and AGN. Blue crosses are quasars classified as point
  sources, and red circles are quasars classified as extended. The
  purple selection line, blue model quasar and red elliptical galaxy
  tracks are as in Fig. \ref{fig:gJK1}, but note that the scale of the
  y-axis has changed.}
  \label{fig:gJK2}
\end{figure}

Fig. \ref{fig:gJK2} shows the location of the confirmed broad-line
quasars and AGN on the $gJK$ diagram, divided by UKIDSS morphological
classification. Blue crosses indicate quasars classified as point
sources, and red circles are extended. As can be seen, the extended
quasars are located between the low redshift end of the model quasar
track and the corresponding low redshift end of the elliptical galaxy
model track, indicating a significant amount of host galaxy flux is
being included in the SDSS PSF and UKIDSS \texttt{aperMag3} magnitude
measurements. The unresolved quasars tend to cluster close to the
model quasar track at higher redshifts. However, there are a
significant number of unresolved quasars located away from the quasar
track, in the region of colour space that is expected to be populated
by red or reddened quasars, as indicated by the black arrow in Fig.
\ref{fig:gJK1}. These objects will be discussed further in Section
\ref{redquasars}.

\begin{figure}
\resizebox{\hsize}{!}{\includegraphics{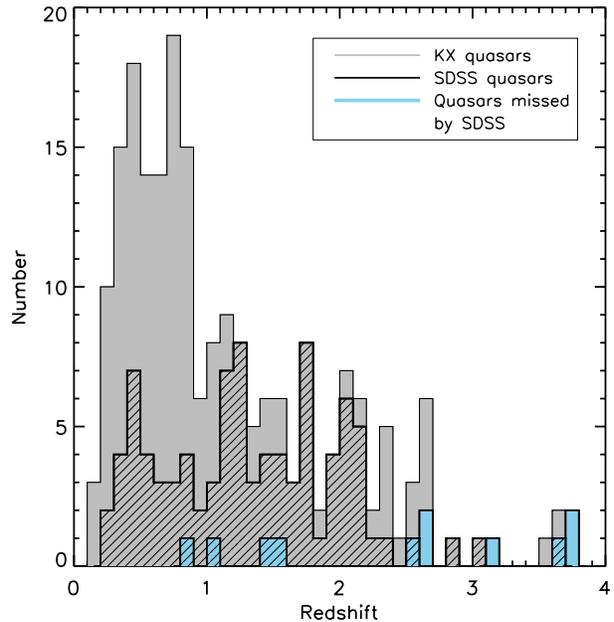}}
\caption{Redshift histogram for confirmed quasars and AGN. The light grey
  histogram contains all broad-line objects identified in this study, and the
  hatched histogram shows recovered SDSS quasars. The blue histogram
  shows quasars that satisfy the SDSS magnitude selection criteria but
  were not targeted for observation by the SDSS quasar selection algorithm.}
\label{fig:nz1}
\end{figure}

Fig. \ref{fig:nz1} shows the redshift distribution of confirmed
quasars and AGN. The solid grey histogram shows all of the broad-line
objects identified in this study, whereas the hatched histogram shows
the SDSS-confirmed quasars recovered. Objects targeted for
spectroscopic observation by the SDSS quasar selection algorithm but
have not yet been observed are not included in the hatched histogram.
The large number of low redshift KX-selected objects highlight the
contribution of host galaxy light in the $K$-band. Most of these objects
have extended morphological classification, and many of the objects at
z $<0.5$ are fainter than the M$_i<-22.4$ magnitude restriction. The
eleven quasars shown in blue are bright enough to have been selected
as SDSS quasar targets but were not in fact included. These objects
are discussed further in Section \ref{Complete}.

\section{Analysis}

We begin by comparing our quasar sample to the SDSS quasar sample in
the same area of sky, dividing the comparison into two redshift
regimes. We address the completeness and the effectiveness of the SDSS
quasar selection algorithm, as well as commenting on the detection of
broad absorption line (BAL) quasars. Next, we consider three different
populations of red objects, contained within our $K$-band limited
sample, possessing mild to severe levels of reddening.


\subsection{Completeness of the SDSS Quasar Catalogue}\label{Complete}

With more than 77\,000 quasars, the SDSS DR5 quasar catalogue
(Schneider et al. 2007) is the largest collection of quasars selected
from a single photometric dataset to date. However, the DR3 quasar
catalogue (Schneider et al. 2005), has been much more thoroughly
studied (Vanden Berk et al. 2005, Richards et al. 2006). Our results
provide an independent test of the completeness of the SDSS quasar
selection. As the SDSS target algorithm differs for z $<3.0$ and z $>3.0$,
the two redshift intervals should be treated separately. For z $<3.0$,
an object must have $i<18.7$ to be selected\footnote{Recall that all
  magnitudes quoted in this work are based on the Vega system}. If the
$ugriz$ colours indicate that the object is probably at z $>3.0$, the
magnitude limit changes to $i<19.8$.

There are 13 KX-selected, spectroscopically confirmed quasars that
have z $<3.0$, $i<18.7$, M$_{i}<-22.4$, and do not have SDSS spectra.
Six of these objects were targeted by the SDSS selection algorithm but
have not been observed.  The remaining seven objects, three of which
are at z $\simeq 2.6$, where the quasar locus comes very close to the
stellar main sequence in optical colour space, eluded the selection
algorithm. The other four objects have redshifts $0.8<$ z $<1.5$.
These seven untargeted objects result in a density of 0.5~deg$^{-2}$
quasars which evade the SDSS selection algorithm, consistent with the
0.44~deg$^{-2}$ found by Vanden Berk et al. (2005).  However, as
Vanden Berk et al. (2005) considered only point sources,
0.44~deg$^{-2}$ is a lower limit. The properties of the seven
untargeted objects found in this sample are very similar to the
population of quasars uncovered by Vanden Berk et al. (2005).
KX-selection successfully identifies as quasar targets objects that
are excluded from the SDSS selection algorithm.

Focusing on the high redshift regime, there are seven KX quasars at
z~$>3.0$, two of which have SDSS spectra, and one of which was
targeted for observation by SDSS. The remaining four quasars include
two objects that have strong BAL troughs. The density on the sky of
KX-selected quasars at $3.0<$~z~$<3.7$ of 0.5~deg$^{-2}$ (7 quasars in
12.8~deg$^2$) compares very well with the SDSS DR3 quasar catalogue
density of 0.48~deg$^{-2}$ (2011 quasars in 4188~deg$^2$).  Therefore,
although there are significantly more z $>3.0$ quasars found in this
study (seven KX-selected compared to three SDSS-selected), the
difference may well be attributed to small number statistics. It
should be noted that the KX selection is extremely effective in
recovering confirmed quasars from the SDSS catalogue. There are 153
SDSS quasars included in our $K\le17.0$ photometric catalogue and all
but one satisfy the KX-selection criteria.


\subsubsection{Broad Absorption Line Quasars}

At least nine objects displaying strong, deep BAL troughs are included
in our KX-selected quasar catalogue. There are three additional
objects that show signs of associated absorption, but the SNR of the
spectra is too low for a definitive BAL identification to be made.
Estimates of the fraction of BAL quasars among the quasar population
range from 10 to 30 per~cent (Chartas 2000, Hewett \& Foltz 2003,
Trump et al.  2006), the difference arising, at least in part, from
difficulties in quantifying the selection of BALs accurately.

The BAL fraction within the SDSS DR3 quasar catalogue has been
computed by Trump et al. (2006). The `classical' BAL population
(objects with balnicity index $>0$) is 10.4$\pm$0.2~per~cent of quasars
at \mbox{1.7~$<$~z~$<$~4.38}. This same fraction holds true for the
more restricted redshift range of \mbox{1.7~$<$~z~$<$~3.8}. For the
KX-selected sample, the percentage of BAL quasars within the entire
population over \mbox{1.7~$<$~z~$<~$3.8} is $\sim$ 15~per~cent (8 BALs
and 44 non-BAL quasars). A recent study from Dai, Shankar, \& Sivakoff
(2007) matches the SDSS DR3 quasar catalogue to 2MASS, and notes an
increasing fraction of BAL quasars with increasingly redder passbands.
Although the Dai et al.  (2007) study uses the much brighter 2MASS
photometry, the result of the BAL fraction increasing at longer
wavelengths appears to be consistent with this small KX-selected
sample. A much larger sample flux-limited in the $K$-band can be
created by matching the SDSS DR3 quasar catalogue to the UKIDSS LAS
DR2+. The fraction of classical BAL quasars at $K<17.0$ and
\mbox{1.7~$<$~z~$<$~3.8} is computed to be 17.5~per~cent, confirming
both the Dai et al.  (2007) result and the result from the present
study. We return to the colour distribution of the BAL quasars in
Section~\ref{discred}.

\begin{figure}
\resizebox{\hsize}{!}{\includegraphics{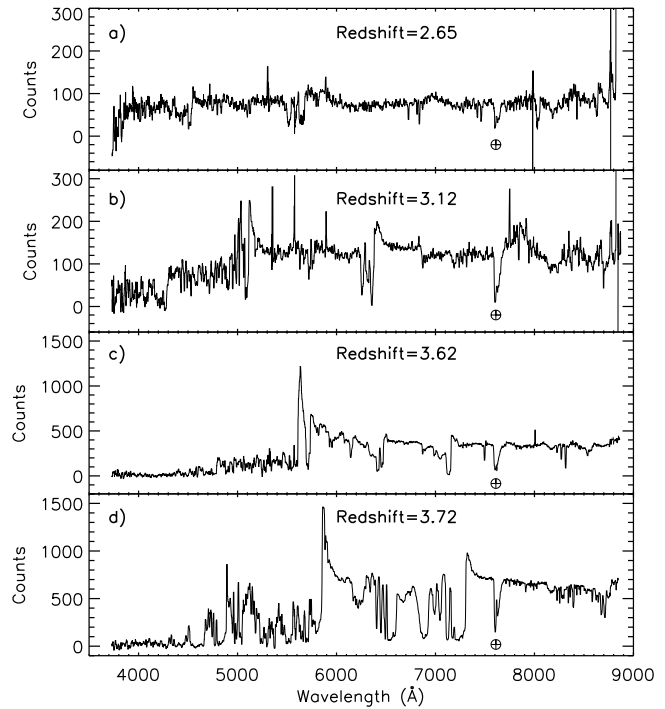}}
\vspace{0.1cm}
\caption{Example spectra of broad absorption line quasars included in
  the sample. The objects' names and redshifts are: a) ULAS
  J142034.43+000451.2, z~$=2.65$; b) ULAS J154258.19+054016.7,
  z~$=3.12$; c) ULAS J130348.94+002010.4, z~$=3.62$; d) ULAS
  J130409.48-000833.9, z~$=3.72$.  Only spectrum c) is included in the
  SDSS quasar catalogue. The absorption feature seen at 7600\,\AA \ is
  the atmospheric A-band.}
\label{fig:bals}
\end{figure}

The BAL quasars lie close to the quasar locus in a $gJK$ diagram, with
the higher redshift objects following the quasar track redward in
$g-J$. The quasar locus comes very close to the selection boundary at
high redshifts.  At z $>3$, absorption associated with Mg\,{\sc ii}
enters into the $J$-band, which could alter their $J-K$ colours such
that they would be moved leftward across the selection boundary. For
the regions where the DR3 quasar catalogue overlaps the UKIDSS LAS
DR2+ area, a cross-match of the BAL catalogue from Trump et al. (2006)
with the DR2+ results in 350 matches, of which 158 possess both
measured $J$ and $K$ magnitudes. Only two of the 158 objects lie to
the left of our KX-selection line, confirming that the KX-selection is
highly effective at identifying BAL quasars. Fig. \ref{fig:bals} shows
spectra of four of the BAL quasars identified in our study. Only one
of the objects shown (spectrum c) is included in the SDSS quasar
catalogue.

\subsection{Red Quasars}\label{redquasars}

As anticipated, the $K$-band flux-limited sample includes a number of
quasars with non-standard SEDs and quasars that are not identified by
the SDSS quasar selection algorithm. Three populations of objects are
discussed here, starting with quasars that are red in $g-J$. Second we
consider objects whose $i-K$ colours are such that their $i$-band
magnitudes are too faint for inclusion in the SDSS quasar catalogue.
Finally, there is a small population of objects with $K\le17.0$, which
are so red that they do not appear in the SDSS photometric catalogue
at all.

\subsubsection{Red in $gJK$}

\begin{figure}
\resizebox{\hsize}{!}{\includegraphics{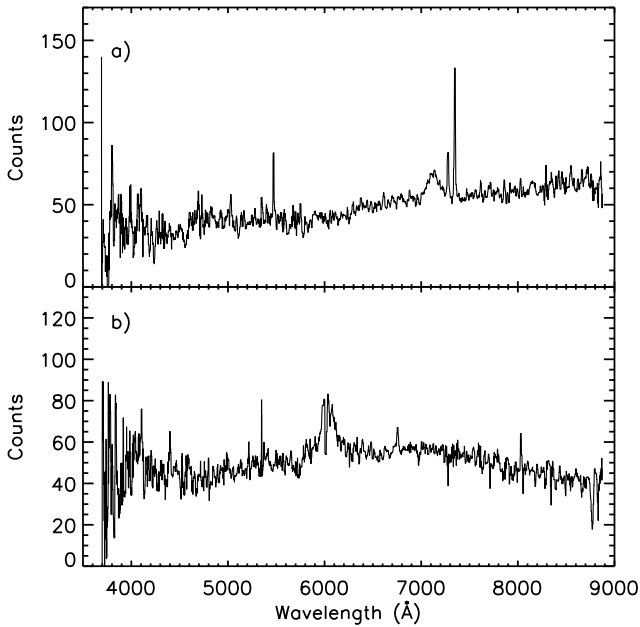}}
\caption{AAOmega spectra of red quasars, corrected as described in the
  text. ULAS J130548+000735 is at z~$=0.47$ and has $g-J=2.81$, while
  ULAS J125438+001447 is at z~$=1.16$ and has $g-J=3.0$. Each spectrum
  has been median filtered with a 15 pixel window.}
\label{fig:redq}
\end{figure}

There are 20 broad-lined objects with $g-J>2.5$, $J-K>1.2$ and M$_i <
-22.0$ in our sample. The spectra for two of the red objects,
corrected for the AAOmega response function, are shown in Fig.
\ref{fig:redq}. Object ULAS J130548+000735, at z~$=0.47$ and
$g-J=2.81$, was selected for spectroscopic follow-up as a
high-redshift candidate by the SDSS. It is classified as extended by
the WFCAM pipeline and the red colours are at least partially due to
host galaxy light. The absolute magnitude in the $i$-band, computed
using a K-correction based on a blue, unobscured, quasar SED, is too
faint for it to be included in the SDSS quasar catalogue, and the host
galaxy flux almost certainly boosted the object into the $K\le17.0$
sample. ULAS J125438+001447, at z~$=1.16$ and $g-J=3$ was not
targeted, as it is fainter than the SDSS quasar catalogue $i$-band
limit. Its stellar morphological classification indicates that the red
colour of this object is most likely due to extinction by dust. This
pair of objects highlights the need for a distinction between objects
that have intrinsically red continua, objects with significant host
galaxy contribution, and objects that are reddened due to the effects
of dust.

As mentioned in Section \ref{quasars}, the dust reddening vector tends
to move objects from the model, unreddened, quasar track into the
region of the $gJK$ diagram populated by galaxies at low redshifts.
While there are 20 objects with $g-J>2.5$, the majority are at
redshift z $\simeq 0.5$, and significant host galaxy contribution
produces their red colours.

\subsubsection{$K\le17.0$, $i>18.7$}

\begin{figure}
\resizebox{\hsize}{!}{\includegraphics{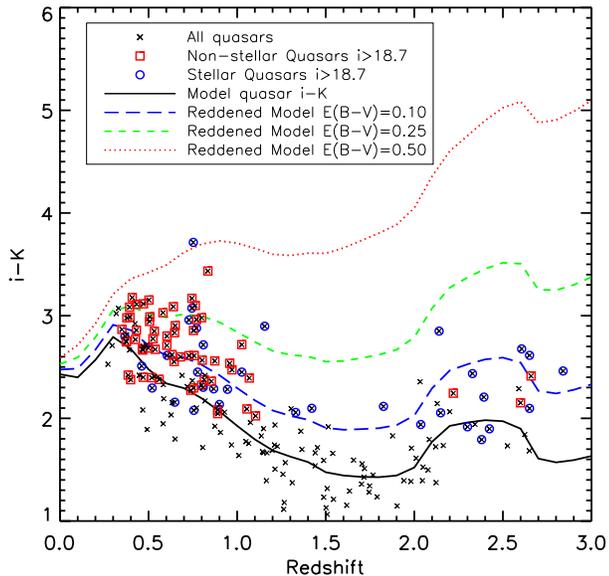}}
\caption{$i-K$ colour of quasars as a function of redshift. The black
  crosses indicate confirmed quasars with $K\le17.0$, M$_i<-22.4$ and
  AAOmega spectra. The blue circles are confirmed quasars with $K\le17.0$,
  M$_i<-22.4$ and $i>18.7$, and are classified as stellar by both
  UKIDSS and SDSS. The red squares are confirmed quasars with $K\le17.0$,
  M$_i<-22.4$ and $i>18.7$, and are classified as non-stellar by
  UKIDSS. The solid black line indicates the $i-K$ colour for a typical
  unobscured quasar, the blue long dashed line indicates the same
  model quasar reddened by SMC-like dust with E($B-V$)=0.10, the green
  short dashed line is the model quasar reddened with E($B-V$)=0.25,
  and the red dotted line is the model quasar reddened with
  E($B-V$)=0.50.}
\label{fig:ifaint}
\end{figure}

There is a significant population of quasars in the KX sample which
are too faint in the $i$-band for inclusion in the SDSS quasar
catalogue. Fig.~\ref{fig:ifaint} shows the population; the black
crosses are all KX-selected quasars with $K\le17.0$, M$_i\le-22.4$, and
z~$\le3.0$, of which there are 189. Of the 189, 93 are fainter than
the SDSS $i$-band limit of $i=18.7$, identified by the open circle and
open square plot symbols. The blue circles mark the 33 quasars which
have been classified as stellar in both the UKIDSS and SDSS
catalogues, whereas the red squares are the 60 quasars classified as
non-stellar in the UKIDSS catalogue. The solid black line indicates
the $i-K$ {\it vs} z colour for the model unobscured quasar, whereas
the blue long dashed line, the green short dashed line and red dotted
line show the $i-K$ colour of the model quasar after being subjected
to SMC-like dust reddening, with E($B-V$) values of 0.10, 0.25, and
0.50, respectively.

As can be seen, the objects that are too faint in the $i$-band are
preferentially redder in $i-K$ than the model quasar, sometimes by
more than a magnitude.  There are three circled objects in
Fig.~\ref{fig:ifaint} that are bluer in $i-K$ than the model quasar at
z $\simeq 2.4$. These can be understood by noting that fixed magnitude
limits in the $i$ and $K$-bands ($i<18.7$ and $K\le17.0$) results in a
constant colour value of $i-K=1.7$.  Due to the presence of the
H$\alpha$ emission line in the $K$-band at z $\simeq2.4$, most quasars
will be redder than $i-K=1.7$, even objects that happen to be bluer
than the model quasar.

The majority of the low redshift objects are classified in the UKIDSS
catalogue as having extended morphology. Host galaxy flux is important
at longer wavelengths and the red colours are in part due to host
galaxy contamination. This effect decreases with increasing redshift
as the quasars become much brighter than the hosts, and thus the red
colours of unresolved objects at z~$\ge 1$ are most likely due to dust
reddening.

There are still a significant number of objects redder than the model
quasar locus that are classified as stellar in both the UKIDSS and
SDSS catalogues, indicating that host galaxy light is not the cause of
their red colours. If dust reddening is responsible, their $i$-band
magnitudes have been underestimated, and are thus excluded from
flux-limited samples defined at optical wavelengths. 

\begin{figure}
\resizebox{\hsize}{!}{\includegraphics{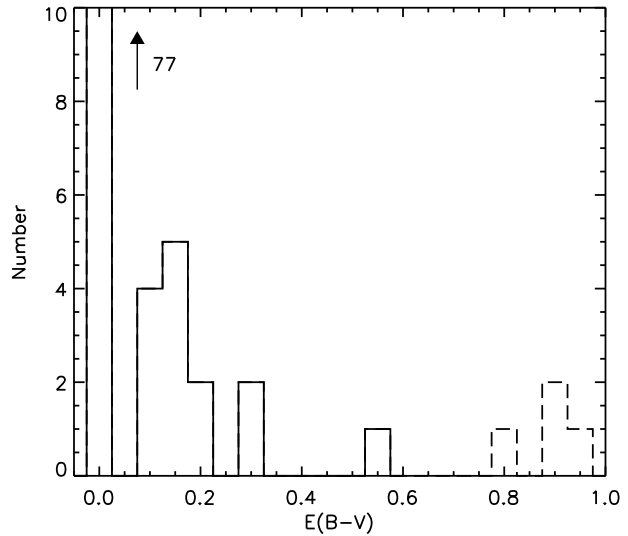}}
\caption{The solid histogram shows the E($B-V$) distribution derived
  from the stellar quasars in Fig.~\ref{fig:ifaint}, while the dashed
  histogram shows the additional values if all four UKIDSS-only
  objects are included at z~$=2$. All 77 quasars lying blueward of the
  \mbox{E($B-V$) $=0.10$} model track are assigned values of
  \mbox{E($B-V$) $=0.0$}.}
\label{fig:ebvhist}
\end{figure}

Assuming that the redder $i-K$ colours are due to dust reddening, the
amount of extinction these objects are suffering can be estimated from
their location on the $i-K$ {\it vs} redshift diagram.  There are 91
stellar, non-BAL quasars with M$_i<-22.4$ in Fig.~\ref{fig:ifaint} for
which the following calculations will be performed. Due to the
apparent intrinsic spread of $i-K$ colours, as displayed by the
distribution of objects with colours both redder and bluer than the
unreddened model quasar seen in Fig.~\ref{fig:ifaint}, objects lying
bluer than the E($B-V$)~$=0.10$ track are assigned \mbox{E($B-V$)
  $=0.0$}. For the 14 stellar, non-BAL quasars redder than
\mbox{E($B-V$) $=0.10$}, SMC-like dust reddening is assumed, and the
amount of extinction in the $K$-band, A($K$), is calculated. The
values range from $0.06<$ A($K$) $<0.41$. A histogram of the estimated
E($B-V$) distribution derived from Fig.~\ref{fig:ifaint} is shown in
Fig.~\ref{fig:ebvhist}. Note that there are 77 objects in the
\mbox{E($B-V$) $=0.0$} bin.

Following a very similar calculation as outlined in Hewett \& Foltz (2003),
for BALs the true number of quasars that would appear in a $K\le17.0$ sample,
correcting for the effects of dust obscuration, can be estimated. For a
population unaffected by dust, the number of objects in a flux-limited
sample is simply the sum of the observed objects, corrected for any
selection effects that may have been introduced. For a population that
is affected by dust, an extra factor is introduced, which accounts for
the fact that the corrected magnitudes are brighter than the nominal
magnitude limit, as:

\begin{equation}\label{eq:sums}
N_{corr}=\sum_{j=1}^{n}\frac{N_{QSO}(K\le17.0)}{N_{QSO}(K\le17.0-A(K)_j)}
\end{equation}

\noindent $N_{corr}$ is the corrected number of objects that should be
observed in a flux-limited sample, and $n=91$ for this KX-selected
sample. For all but 14 of the objects in the sum, the fraction will be
unity, as their corresponding A($K$) values are zero. The corrected
number of quasars that should be observed to $K\le17.0$, accounting
for the dust reddening derived from Fig.~\ref{fig:ifaint}, is
$N_{corr}=97$.  This implies that only six~per~cent ($97-91/97$) of
the total population is missing from this sample due to obscuration.
In an effort to determine how sensitive this result is to small
changes in the E($B-V$) distribution, the same calculations are
performed with the quasar with E($B-V$) $>0.5$ at z~$\simeq0.7$
excluded. The resulting changes in the computed missing fractions are
of the order of one~per~cent.

A similar calculation can be performed for the SDSS DR3 quasar
catalogue, flux-limited in the $i$-band and restricted to z $\le3.0$, using
the A($i$) values derived from the E($B-V$) distribution of the
KX-selected sample.  Estimates of the fraction of the total population
that is missing from the observed $i$-band selected sample reach
the much larger value of $\sim$~30~per~cent.  

If any of the four UKIDSS-only detections turn out to be quasars, the
fraction of obscured objects increases significantly for both $K$-band
and $i$-band samples. We consider the identification of the four
objects next. .

\subsubsection{UKIDSS-only Detections}\label{nosdss}

\begin{figure}
\begin{center}
\resizebox{5cm}{!}{\includegraphics{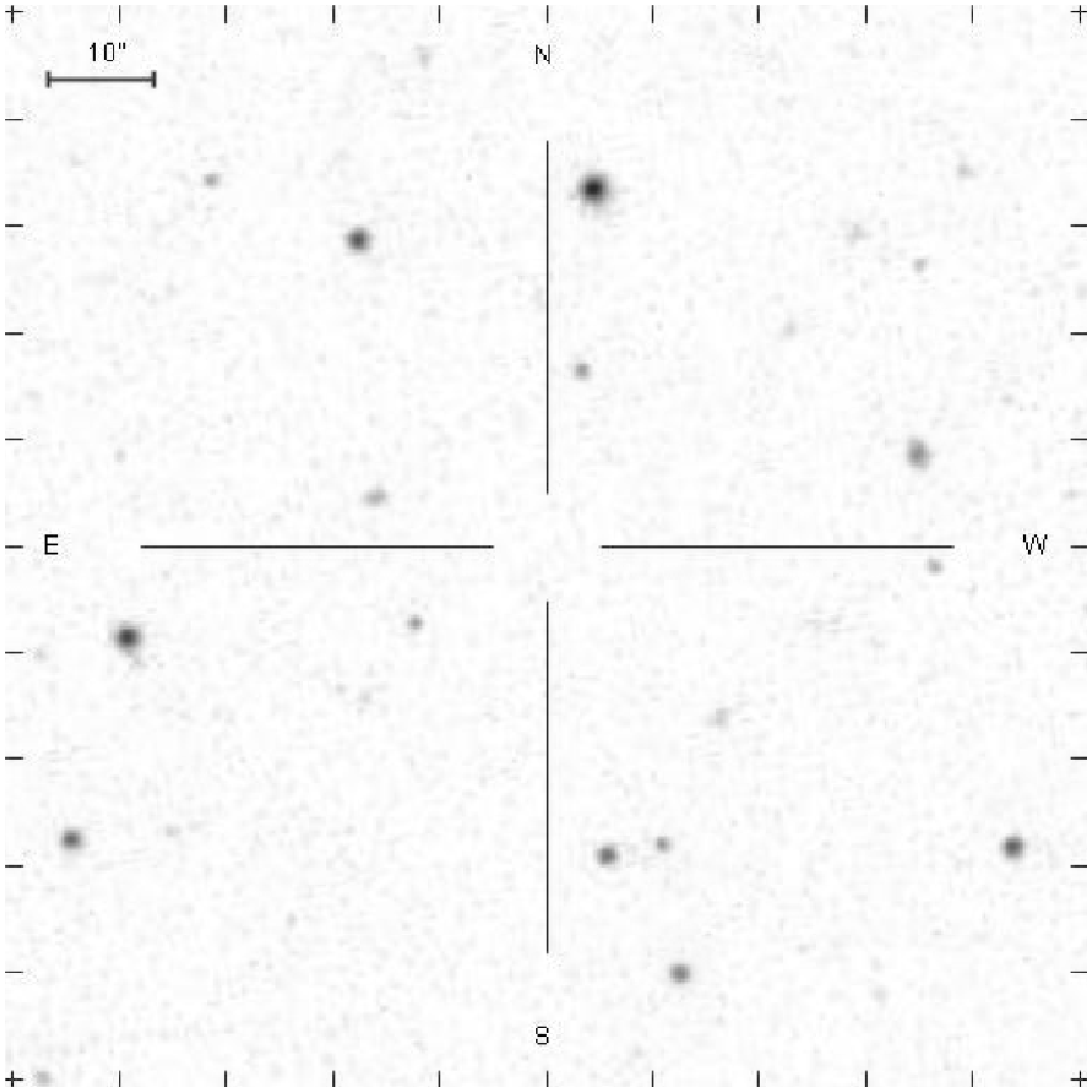}}
\resizebox{5cm}{!}{\includegraphics{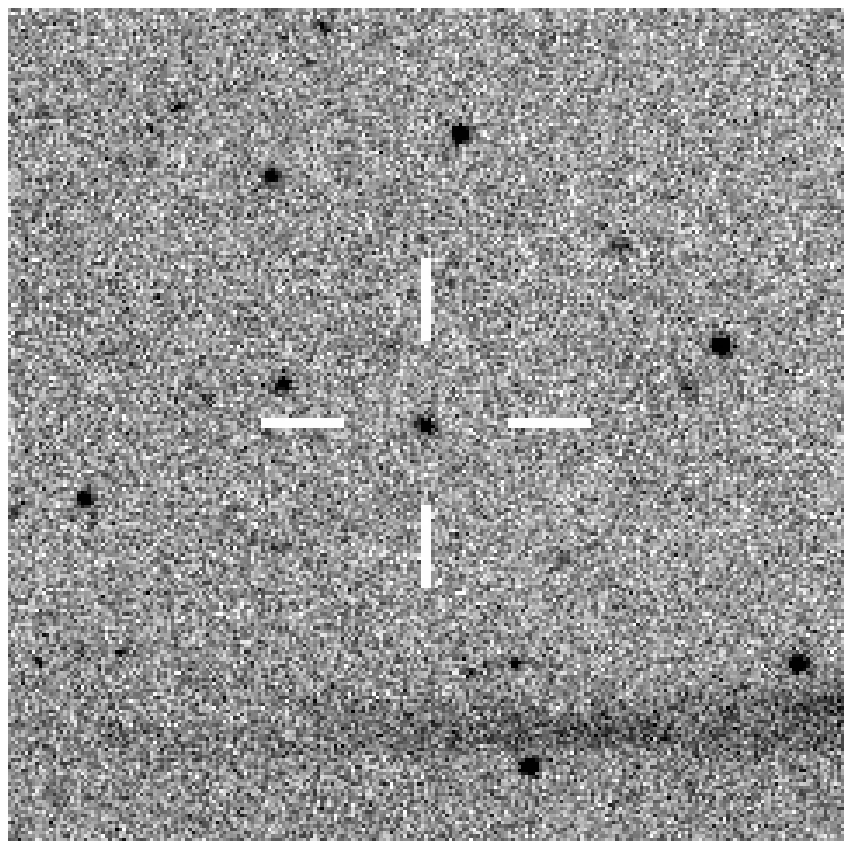}}
\caption{(Top) Composite colour SDSS image for one of the four objects
  detected in the UKIDSS bands and not in the SDSS.  (Bottom) WFCAM
  $K$-band image of the same object, showing a clear detection in the
  UKIDSS bands and not in the SDSS. The scale of both images is shown
  by the 10\arcsec \ bar in the top left corner of the SDSS image.}
\label{fig:nosdss_s}
\end{center}
\end{figure}

Six objects that were detected in the UKIDSS bands and not in SDSS
were observed spectroscopically, which resulted in two
identifications. Objects listed as ULAS J125254-000947 and ULAS
J131049-001514 were both classified as absorption line galaxies. The
other four spectra did not have sufficient flux for identifications to
be made. Image cutouts of one of the four unclassified objects are
shown in Fig. \ref{fig:nosdss_s}.  The $ugrizYJHK$ magnitudes and
magnitude limits are listed in Table \ref{tab:nosdss}, along with
their $i-K$ colours. The calculated E($B-V$) required in order for
each object to have $i-K$ colours consistent with that of the
unreddened  model quasar are also listed, assuming the objects are at 
z $=1$ or z $=2$.

The NIR-extended Photoz can be used to determine which object SEDs are
consistent with the colours of the UKIDSS-only objects. For the two
spectroscopically classified UKIDSS-only objects, the NIR-extended
Photoz produces consistent fits to galaxy SEDs at redshifts similar to
the spectroscopically determined redshifts, lending support to the
spectroscopic classifications. The photometry of ULAS J125946-001135
is consistent with both an emission line galaxy at z$\sim$2.15 and a
highly reddened quasar at z$\sim$0.7. Using the measured $i$-band
magnitude and a K-correction for an Sc-type galaxy, an emission line
galaxy at z$\sim$2.15 would be unfeasibly bright at M$_i\sim-27$,
reducing the confidence in the galaxy classification. ULAS
J131910+000956 is similarly consistent with an emission line galaxy at
z$\sim$1.65 and also a highly reddened quasar at z$\sim$2.1, but also
an L dwarf star of spectral class L3. Again, an emission line galaxy
at z$\sim$1.65 would be unphysically bright, making the galaxy
classification the least likely of the three. ULAS J154727+052451 is
somewhat consistent with an absorption line galaxy at z$\sim$1.7, but
a much better fit to a class L4.5 L dwarf. The best fit to ULAS
J154827+054821 is an Sb-type galaxy at z$\sim$2.6, with the very red
$JHK$ colours difficult to fit with any of the models. Further
observations, preferably NIR spectra, are required in order to
identify these four objects unambiguously.  ULAS J131910+000956 has a
NIR spectrum, but no features have been identified (R G McMahon,
private communication). The implications for these four objects in the
context of a putative population of highly obscured quasars is
discussed in Section \ref{discred}.

\begin{table*}
  \caption{Measured magnitudes and magnitude limits for the six
    UKIDSS-only detections. The few detections in the SDSS bands result from 
    re-analysis of the SDSS data (B. Venemans, private communication).}
\label{tab:nosdss}
\centering
\scriptsize{
\begin{tabular}{ccccccccccccc}\\ \hline

Object & $u$ & $g$ & $r$ & $i$ & $z$ & $Y$ & $J$ & $H$ & $K$ & $i-K$ &
E($B-V$) & E($B-V$) \\ 
Name & & & & & & & & & & & (z = 1) & (z = 2) \\ \hline
ULAS J125254.31-000947.1 & $>$21.9 & $>$23.9 & $>$23.4 & 21.1 & 20.0 &
19.8 & 18.9 & 18.1 & 16.8 & 4.3 & --- & --- \\
ULAS J125946.54-001135.8 & $>$21.8 & $>$23.8 & $>$23.1 & 21.1 & 20.6 & 
19.9 & 19.3 & 18.1 & 16.6 & 4.5 & 1.12 & 0.81 \\ 
ULAS J131049.35-001514.5 & $>$21.9 & $>$23.9 & 21.8 & 21.3 & 20.0 &
19.1 & 18.6 & 18.1 & 17.0 & 4.3 & --- & --- \\
ULAS J131910.65+000956.1 & $>$21.8 & $>$23.3 & $>$23.4 & $>$22.2 & $>$20.5
& 20.2 & 18.7 & 17.2 & 16.0 & $>$6.2 & 1.34 & 0.94 \\ 
ULAS J154727.87+052451.9 & $>$21.6 & $>$24.0 & $>$23.4 & $>$22.8 & $>$20.7
& 20.0 & 18.5 & 17.6 & 16.8 & $>$6.1 & 1.30 & 0.92 \\
ULAS J154827.70+054821.5 & $>$21.8 & $>$24.1 & $>$23.4 & $>$22.4 & $>$20.4 & 
$>$20.8 & 20.2 & 17.7 & 16.4 & $>$6.0 & 1.26 & 0.89 \\ \hline
\end{tabular}
}
\end{table*}


\section{Discussion}

The population of quasars selected in the KX sample that do not appear
in the SDSS catalogue is briefly discussed here, along with the
implications for uncovering a population of highly reddened quasars.
The observations are compared to simulations based on a quasar
luminosity function defined in the $b_J$-band, and then compared to
results from recent studies.


\subsection{A Population of Red Quasars}\label{discred}

\subsubsection{Mild Reddening}

Selecting quasars in the $K$-band has been shown to be very effective
at finding not only standard, blue quasars, but also quasars with red
SEDs, either due to red intrinsic SEDs, host galaxy light, dust
reddening or the presence of BALs.

The distribution of $i-K$ colours as a function of redshift is shown
in Fig.~\ref{fig:ifaint}. The SDSS quasar catalogue can be used to
provide a comparison sample. Cross-matching the DR3 quasar catalogue
to the UKIDSS DR2+ catalogue results in 3401 matches, 969 of which
are stellar, have measured $K$-band magnitudes $K\le17.0$ and are at z
$\le3.0$. Extended objects are excluded from this comparison as it is
understood that the $K$-band selected sample is more sensitive to host
galaxy flux than the optically selected sample. The comparison is also
restricted to z $\le3.0$ as the SDSS $i$-band faint magnitude limit
changes at this redshift.

Another population to be considered separately are the BAL quasars.
Extracting the matched SDSS DR3 -- UKIDSS DR2+ quasars at
\mbox{1.7~$<$~z~$<$~3.8}, and dividing the sample into BAL (144) and
non-BAL (249) quasars, a two-sided Kolmogorov-Smirnov (K-S) test
comparing the $i-K$ distributions of each sub-population returns the
probability that the two sub-populations are drawn from the same
parent population as 5$\times$10$^{-10}$.  The BAL quasars are
preferentially redder in $i-K$.  The cause of the reddening is not
known definitely but the presence of an additional average value of
E($B-V$)=0.02--0.03 affecting the BAL quasars, consistent with the
findings of Reichard et al.  (2003) and Dai et al.  (2007), can
explain the colour difference.

After removing the extended objects and BAL quasars, the $i-K$
distributions for the remaining stellar objects can be compared.  A
K-S test comparing the SDSS DR3 stellar quasars and the KX-selected
stellar quasars for \mbox{0.0~$<$~z~$<$~3.0}, excluding the BAL
quasars, returns a probability of 0.45 that they are from the same
underlying distribution.  Although there is no evidence for a
difference in the $i-K$ distributions for the samples as a whole,
there is an indication of the presence of a red `tail' in the
KX-selected sample.  Specifically, of 830 stellar SDSS quasars between
\mbox{0.0~$<$~z~$<$~3.0}, 65 (8~per~cent) are redder than the model
quasar reddened by E($B-V$)=0.10, compared to 14 of 87 (16~per~cent)
within the KX-selected sample.  Eight of 830 (1~per~cent) SDSS quasars
are redder than the model reddened by E($B-V$)=0.25, compared to 3 of
87 (3~per~cent) of the KX-selected sample. Using a binomial statistic,
the probability of finding 14 of 87 BALs at E($B-V$)~$>0.1$ if they
share the same $i-K$ distribution as the non-BALs is only
0.7~per~cent. Similarly, there is a 5~per~cent chance of finding 3 of
87 BALs at E($B-V$)~$>0.25$ if the BALs and non-BALs have the same
$i-K$ distribution.

Note that although we have identified a number of quasars with red
$i-K$ colours, there does seem to be an upper limit to the
distribution.  From Fig.~\ref{fig:ifaint}, it is apparent that there
are no confirmed quasars with $i-K>4.3$, even though the experiment is
sensitive to objects with redder colours, as demonstrated by the fact
that galaxies with $i-K>4$ are successfully identified.

\subsubsection{Severe Reddening}

It is very important to secure identifications for the four
unclassified objects in Section \ref{nosdss}. As seen in Table
\ref{tab:nosdss}, three of the four have $i-K>6$. If these objects are
quasars, they will lie two magnitudes redder in $i-K$ than the
reddest quasars in Fig. \ref{fig:ifaint}, leaving a large gap in $i-K$
that is seemingly unpopulated. Physical models consistent with such a
distribution would include a bimodal distribution of E($B-V$), or a
very strong dependence of severe reddening on the intrinsic luminosity
of the quasars.

Using the observed $i-K$ values listed in Table \ref{tab:nosdss} as
well as the model quasar colours, estimates of each object's
unreddened $i$ and $K$-band magnitudes can be made. If the four
objects are quasars at z $=1$ they would have intrinsic colours of
$i-K\simeq1.93$ and would be experiencing
\mbox{$0.7<$~E($B-V$)~$<1.2$} of reddening. Corresponding unreddened
magnitudes fall in the range $14.9<K<15.9$, putting them at the bright
end of the quasar magnitude distribution. Adopting a redshift of z$=2$
instead, their unreddened magnitudes fall in the range $14.2<K<15.4$,
with the brightest of the four brighter than all the quasars included in
this study.


\subsection{Comparison with Previous Predictions}

In an effort to put new NIR observations in context with the existing
optical data, simulated surveys in a variety of passbands, including
the UKIDSS $K$-band, were performed by Maddox \& Hewett (2006).
Briefly, the simulations were based on a quasar luminosity function
(QLF) derived from observations spanning $16<b_J<20.85$ by Croom et
al.  (2004), and a quasar SED that matched the median colours of
bright SDSS DR3 quasars for $0.1<$ z $<3.6$. Host galaxy brightnesses
were estimated from the $g-r$ colours of low redshift SDSS quasars and
were added to that of the quasars according to three different
relationships between the luminosity of the quasar and the luminosity
of its host galaxy.  The additional host galaxy flux serves to boost
the magnitude of the combined quasar + host galaxy system, and alters
the shape of the combined SED. The combined quasar + host galaxy SED
is used to convert the QLF from the $b_J$ passband to the passband of
choice, resulting in number-magnitude and number-redshift relations
for the chosen passband.

\begin{figure}
\resizebox{\hsize}{!}{\includegraphics{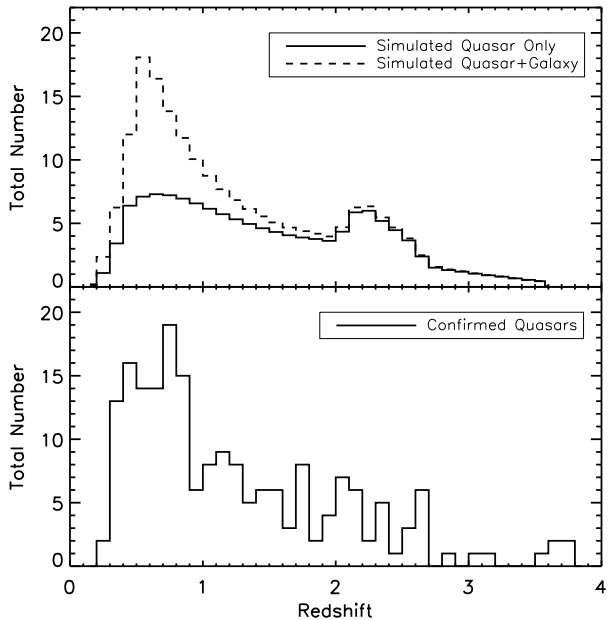}}
\caption{Comparison of simulated (top panel) and observed (bottom
  panel) number-redshift histograms for the 12.8~deg$^2$, $14.0<K<17.0$
  survey. The solid black line in the top panel considers light from the
  quasars only, while the red dashed line considers light from both
  the quasar and the host galaxy. The simulations are described in detail
  by Maddox \& Hewett 2006.}
\label{fig:nz2}
\end{figure}

Fig. \ref{fig:nz2} shows a comparison of the results from the
simulations, using the updated quasar SED described in Section
\ref{qcandidates}, with the results of the 12.8~deg$^2$ of
observations presented here. The solid line in the top panel considers
the results for a simulated survey for $14.0<K<17.0$ and 12.8~deg$^2$
including only flux received from the quasars, while the dashed line
includes light from both the quasars and their host galaxies. As can
be seen, host galaxy light at long wavelengths makes a significant
contribution to the total light, and affects the results to z $\simeq
2$. The solid line in the bottom panel is the actual redshift
distribution of the observed quasars brighter than M$_i<-22.4$ in this
study. The agreement in overall shape and normalisation is very good,
except at very high redshifts, as the simulations were limited to z
$\le3.6$.

From this comparison, it becomes apparent that the importance of the
additional light contributed by the quasar host galaxies at longer
wavelengths has been successfully accounted for, as indicated by the
large population of morphologically extended objects selected in the
$K$-band that do not appear in the $i$-band selected sample, and the
good agreement between the observed and simulated number-redshift
relations. In converting the QLF defined in the $b_J$-band to the
$K$-band, only additional host galaxy flux was added, with no
corrections or additions made based on a population of dust-reddened
objects that would be visible in the $K$-band but not in the
$b_J$-band. A large, moderately obscured population would be apparent
in the comparison as a large excess in the $K$-band selected sample
not present in the simulated results, which is not seen in
Fig~\ref{fig:nz2}. This contribution from the quasar hosts at longer
wavelengths and low redshifts complicates the construction of
luminosity functions and magnitude distributions, for example, due to
the difficulty of accurately separating the quasar from the host
galaxy light.

\subsection{Comparison with Recent Studies}

Work has recently been published by Glikman et al. (2007) studying
radio selected quasars with very red ($R-K>4$) colours over a large
area of sky. Unfortunately, as their NIR photometry is taken from the
2MASS survey, the magnitude range for which their survey is complete
($K<14$) is too bright to overlap with this study. However, the
simulations described above can be used to supplement the present
study at brighter magnitudes.

Glikman et al. (2007) conclude that, based on their observations, red
quasars add an extra 25-60~per~cent to the unreddened quasar
population at $K\le14.0$. This claim can be tested by using a
combination of the spectroscopic results and the simulations. From the
simulations, we expect 0.081 unreddened quasars deg$^{-2}$ for
$K\le14.0$. If a further 60~per~cent of the total population are red,
one expects an additional 0.122 red quasars deg$^{-2}$ to appear at
fainter $K$-band magnitudes. Due to our small survey area, this
density of red quasars would add only an additional one or two objects
to our sample, and we cannot make a strong statement regarding their
fraction with respect to the unreddened population;  $\sim$150 deg$^2$
of area would be required in order to find 10 unreddened quasars with
$K<14.0$.

However, assuming the 60~per~cent fraction does not depend on
magnitude, the same procedure can be followed for fainter magnitudes.
For $14<K<15$, the simulations predict 0.46 unreddened quasars
deg$^{-2}$, or six quasars for our effective area of 12.8 deg$^2$. If
an additional 60~per~cent of the entire population is red, then we
expect an extra 0.69 red quasars deg$^{-2}$, or nine red quasars to
appear at fainter magnitudes. Even if all four of the UKIDSS-only
detected objects are heavily reddened quasars, they would only account
for half of the predicted number.

The number of red quasars increases at even fainter magnitudes, as the
unreddened population increases significantly. For $15<K<16$, one
expects an extra 49 quasars to appear at $K>16$. As seen from columns
2 and 3 of Table \ref{tab:glik}, the discrepancy between the simulated
and observed numbers of quasars in each magnitude interval is too
small to harbour such large numbers of excess quasars, as listed in
column 4.

If, instead, we adopt the obscured fraction to be their quoted lower
limit of 20 per cent, one expects fewer extra red quasars, as listed
in column 5 of Table \ref{tab:glik}. Based on the small differences
between the simulated and observed numbers of quasars in each
magnitude range, our results strongly favour an obscured quasar
fraction of $\la$20~per~cent.

A similar study searching for rare, highly reddened objects is being
undertaken by Hawthorn et al. (2008, in preparation), using
100~deg$^2$ of UKIDSS LAS data. 22 candidates with $K\le17.0$ and
$J-K>2.5$ are selected for further NIR spectroscopic observation. Of
11 candidates observed, seven are clearly identified as broad-lined
quasars, while the other four did not produce a definitive
identification. One of the four unidentified objects is ULAS
J131910+000956, listed in Table~\ref{tab:nosdss}. The seven
identifications result in a lower limit on the surface density of
0.14~deg$^{-2}$. Thus one or two red objects are expected in the
12.8~deg$^2$ of the current study, consistent with one or two of the
objects listed in Table~\ref{tab:nosdss} being quasars.

\begin{table}
\centering
\caption{\label{tab:glik} Simulated, observed and proposed numbers of missing quasars for different magnitude ranges within 12.8 deg$^2$. The numbers in the second column are derived from the simulations described in the text, while the numbers in the third column are directly from the spectroscopic observations. The fourth and fifth columns contain the number of red quasars that would be missing due to dust obscuration if the obscured fraction is 60~per~cent and 20~per~cent, respectively. The small differences between the simulated and observed numbers of quasars does not favour a large fraction of obscured objects.}
\begin{tabular}{ccccc} \hline
Magnitude & 12.8 deg$^{-2}$ & 12.8 deg$^{-2}$ & Red Q & Red Q \\ 
Range & Sim & Obs & 60\% & 20\% \\ \hline
$13<K<14$ & 1.037 & --- & 1 & --- \\
$14<K<15$ & 5.850 & 3 & 9 & 1 \\ 
$15<K<16$ & 32.99 & 33 & 49 & 8 \\ 
$16<K<17$ & 149.06 & 160 & 224 & 37 \\ \hline
\end{tabular}\\
\end{table}

A recent study by Jurek et al. (2007), employing a complete
spectroscopic sample derived from the Fornax Cluster Spectroscopic
Survey, tests the effectiveness of a variation of KX selection at
detecting quasars. Instead of using $g-J$ and $J-K$, the combination
of colours $b_J-R$ and $R-K$ is shown to be at least as effective as
optical two colour selection, with less bias against dust reddened
objects. Although this combination of an optical and optical--NIR
colour shows the same property that dust reddened quasars remain
separate from the stellar locus, there is significant overlap of the
blue quasars and stars in colour space, which is not present when
using an optical--NIR and a NIR colour. As seen in
Fig.~\ref{fig:gJK1}, the NIR $J-K$ colour provides the separation of quasars
from stars, highlighting the importance of having high-quality NIR
data in at least two different passbands.


\subsection{Implications for Future Near-Infrared Surveys}

As this was the first large-area KX-selected quasar survey performed
to date, the goal of the study was to undertake a comprehensive survey
without regard to efficiency. The single cut in $gJK$ colour space
succeeded in selecting more than 99~per~cent of the SDSS quasars with
$K\le17.0$ within the survey area, and has thus been shown to be
highly effective in recovering known quasars. The analysis in
Section~\ref{Complete} also demonstrates the ability of the
optical-NIR colour selection to include quasars missed by the SDSS
optical colour selection algorithm.

The results from this initial sample may now be used as input for
future observations, to improve the observing efficiency. The use of
photometric redshifts to identify low redshift (z~$\le0.8$) inactive
galaxies, together with the radial light profile information described
in Section~\ref{exflux}, will greatly improve the efficiency of the KX
selection by reducing the number of contaminants included as quasar
candidates. Particular subclasses of objects can also be more
efficiently targeted with the extra information provided by their
location on the $gJK$ diagram. High redshift BALs tend to be confined
to $J-K=1$, $g-J>2.5$, and highly reddened quasars at z $>1$ will be
point sources located amongst the cloud of morphologically extended
galaxies.


\section{Summary}

With nearly 200 quasars over an effective area of 12.8~deg$^2$, this
is the largest $K$-band selected quasar sample to date by a factor of
twenty, utilising the high quality, large area near-infrared data
provided by UKIDSS.  The KX-selection, exploiting the $K$-band excess
of quasars with respect to stars, is equally sensitive to both
standard blue quasars and dust-reddened objects over the entire
redshift range available.  A number of genuine quasars at z $<3$ not
initially selected by the multicolour technique employed by the SDSS
are included, consistent with the completeness calculations performed
by Vanden Berk et al.  (2005).  More than twice as many high redshift
quasars are found with this selection than are targeted by the SDSS,
however the difference can be attributed to small number statistics.
The selection is also sensitive to quasars with unusual SED shapes,
such as BAL quasars.

Using the $ugrizYJHK$ photometry as well as morphological information,
virtually all of the candidates that were not observed
spectroscopically have been classified using the SDSS Photoz algorithm
and a new NIR-extended photometric redshift scheme. The population of
objects consistent with both galaxy and optically obscured quasar SEDs
harbouring nuclear components bright enough for inclusion in this
$K$-band flux-limited sample has been ruled by using the $i$-band
radial light profile information along with $K$-band flux ratios.

The distribution of $i-K$ colours as a function of redshift is compared to that
from the optically-selected SDSS DR5 quasar catalogue, and is found to include
more objects with redder $i-K$.  The observed BAL quasars are preferentially
redder than the bulk of the quasar population, the BALs possessing colours
consistent with an additional average of E($B-V$)=0.02--0.03 of SMC-type dust
reddening.

The fraction of quasars missing from this $K$-band selected sample due
to dust reddening is computed to be $<$10~per~cent, whereas the
fraction missing from a sample selected in the $i$-band is
considerably larger at $\sim$30~per~cent. Four objects, detected in
the NIR but not in the SDSS optical bands, remain unclassified. Their
identification is essential, as if all four are quasars, the heavily
dust-reddened fraction of the most luminous quasars would be very large.

With $ugrizYJHK$ photometry, morphology classification, and spectra
for more than 3000 objects, this dataset contains a wealth of
information which is yet to be fully exploited.  The results from this
initial sample can be used as input for future observations, improving
the observing efficiency by reducing the number of contaminants.
Particular subclasses of quasars can also be more efficiently targeted
using their location in the $gJK$ diagram, with high redshift BALs
confined to $J-K=1$, $g-J>2.5$, and highly reddened quasars at z $>1$
will be point sources located amongst the cloud of morphologically
extended galaxies.  This modest sized sample has already placed
improved constraints on the fraction of obscured quasars missing from
optically-selected samples, and as larger areas are surveyed, this
fraction can be constrained even further.


\section*{Acknowledgments}

We acknowledge the contributions of the staff of UKIRT to the
implementation UKIDSS survey and the Cambridge Astronomical Survey
Unit and the Wide Field Astronomy Unit in Edinburgh for processing the
UKIDSS data. This work is based in part on data obtained as part of
the UKIRT Infrared Deep Sky Survey. The United Kingdom Infrared
Telescope is operated by the Joint Astronomy Centre on behalf of the
UK Particle Physics and Astronomy Research Council.

This work was based in part on observations made with the
Anglo-Australian Telescope. We warmly thank all the present and former
staff of the Anglo-Australian Observatory for their work in building
and operating the 2dF and AAOmega facilities.

Funding for the Sloan Digital Sky Survey (SDSS) has been provided by
the Alfred P. Sloan Foundation, the Participating Institutions, the
National Aeronautics and Space Administration, the National Science
Foundation, the U.S. Department of Energy, the Japanese
Monbukagakusho, and the Max Planck Society. The SDSS Web site is
http://www.sdss.org/.

The SDSS is managed by the Astrophysical Research Consortium (ARC) for
the Participating Institutions. The Participating Institutions are The
University of Chicago, Fermilab, the Institute for Advanced Study, the
Japan Participation Group, The Johns Hopkins University, Los Alamos
National Laboratory, the Max-Planck-Institute for Astronomy (MPIA),
the Max-Planck-Institute for Astrophysics (MPA), New Mexico State
University, University of Pittsburgh, Princeton University, the United
States Naval Observatory, and the University of Washington.

This publication makes use of data products from the Two Micron All
Sky Survey, which is a joint project of the University of
Massachusetts and the Infrared Processing and Analysis
Center/California Institute of Technology, funded by the National
Aeronautics and Space Administration and the National Science
Foundation.

We thank the anonymous referee for improving the presentation of this
paper. NM wishes to thank the Overseas Research Students Awards
Scheme, the Cambridge Commonwealth Trust, and the Dr. John Taylor
Scholarship from Corpus Christi College for their generous support.


\appendix

\section{Photometric Redshifts}\label{appb}

\begin{figure}
\resizebox{\hsize}{!}{\includegraphics{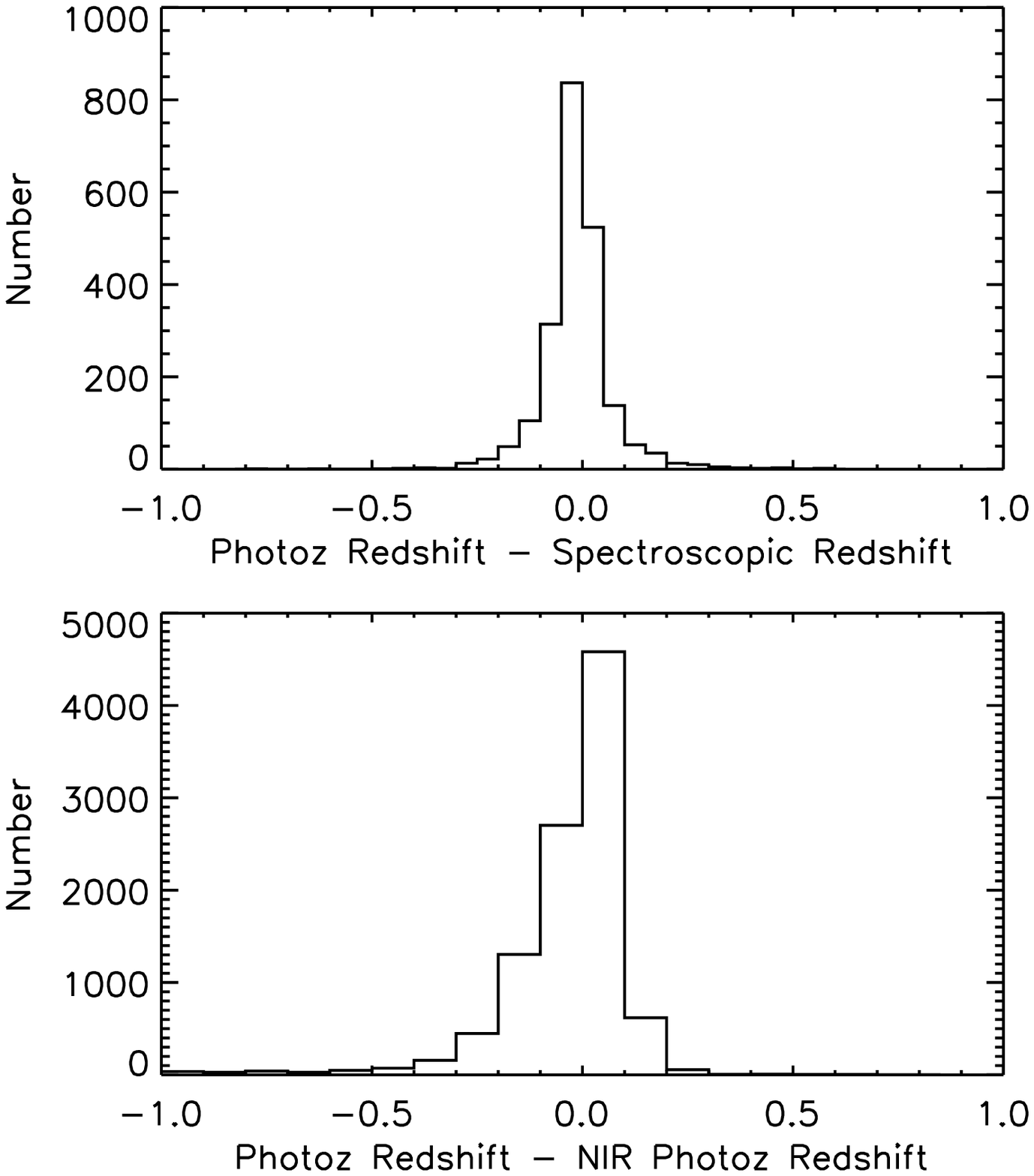}}
\caption{(Top panel) The difference between the computed SDSS Photoz
  redshift and the spectroscopically determined redshift for confirmed
  galaxies with confident Photoz results. (Bottom panel) The
  difference between the SDSS Photoz and the NIR-extended Photoz
  redshifts, for the entire population of extended candidates for
  which confident results exist for each.}
\label{fig:pz1}
\end{figure}

Two photometric redshift routines were employed to classify the many
thousands of extended objects within the survey area that were not
observed spectroscopically, and the few objects with spectra that
remained unclassified. The SDSS Photoz routine is optimised for use
with galaxies, whereas the NIR-extended Photoz is capable of
identifying unresolved objects as well. The top panel of
Fig.~\ref{fig:pz1} shows the difference between the SDSS Photoz
redshift and the spectroscopic redshift for spectroscopically
classified galaxies that possess confident Photoz results. The bottom
panel of Fig.~\ref{fig:pz1} compares the SDSS Photoz redshift to the
NIR-extended Photoz redshift, for extended candidates that possess
confident results from both routines. The two sets of photometric
redshifts are highly consistent with spectroscopic redshifts and with
each other.

\begin{figure}
\resizebox{\hsize}{!}{\includegraphics{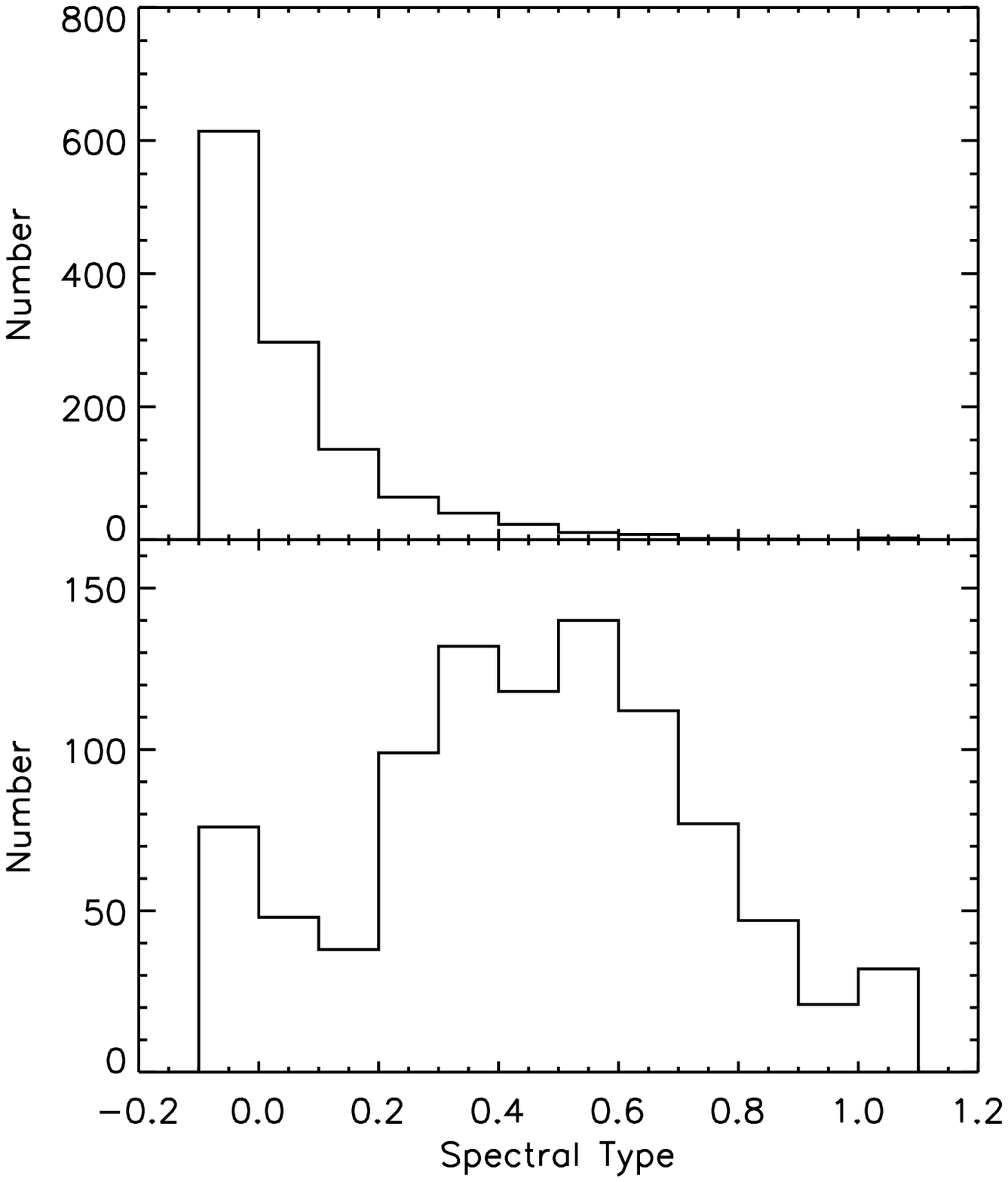}}
\caption{The distribution of spectral types as determined by the SDSS
  Photoz algorithm for the spectroscopically classified absorption
  line galaxies (top panel) and star-forming galaxies (bottom panel).}
\label{fig:pz2}
\end{figure}

Fig.~\ref{fig:pz2} displays the ability of the SDSS Photoz routine to
classify the spectral type of an extended object as well as fitting
the redshift. The top panel shows the resulting spectral type
distribution for objects spectroscopically classified as absorption
line galaxies, with values near zero indicating old, passively
evolving stellar populations. The bottom panel of Fig.~\ref{fig:pz2}
shows the resulting spectral types for spectroscopically classified
emission line galaxies, with values greater than zero indicating
increasing amounts of star formation.

\begin{figure}
\resizebox{\hsize}{!}{\includegraphics{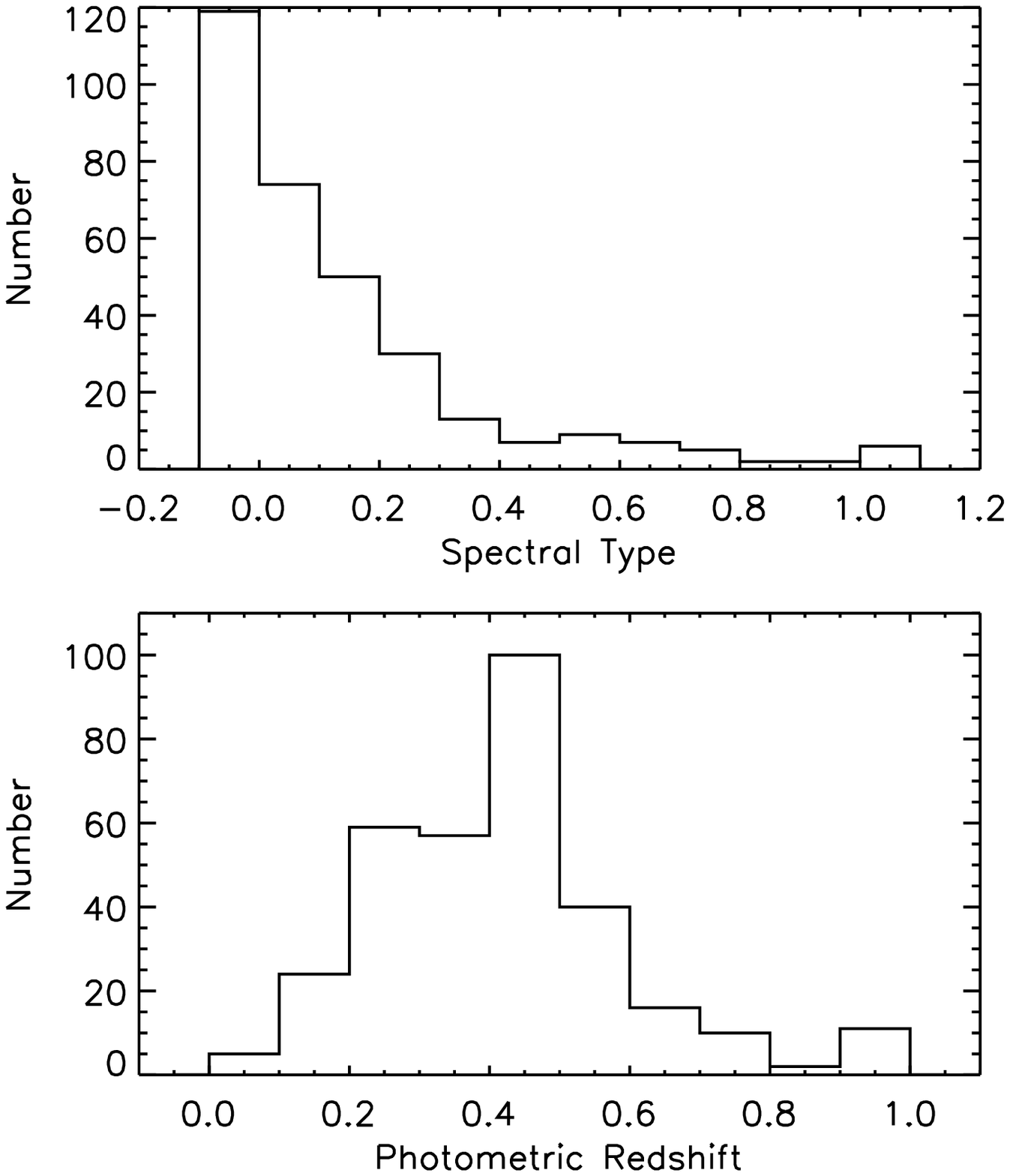}}
\caption{The distribution of spectral types as determined by the SDSS
  Photoz algorithm for the unclassified objects (top panel), and their
  resulting photometric redshifts (bottom panel).}
\label{fig:pz3}
\end{figure}

From the results shown in Fig.~\ref{fig:pz1} and Fig.~\ref{fig:pz2},
confidence in the SDSS Photoz output is gained. Fig.~\ref{fig:pz3}
shows the SDSS Photoz results for the 322 objects with spectra but no
classification, and confident Photoz results. The top panel shows that
this population is dominated by absorption line galaxies (spectral
type near zero), with a small population of objects with moderate star
formation, consistent with the conclusions drawn from
Fig.~\ref{fig:noid}. The bottom panel of Fig.~\ref{fig:pz3} puts the
bulk of the population of objects at z~$\simeq0.4$.

\section{Sample Composition: Galaxies and Stars}\label{appa}

As objects with extended morphology were not discriminated against
during selection of candidates for follow-up spectroscopy, there are a
significant number of galaxies included in the spectroscopic sample. 
Type 2 Seyfert galaxies are classified by the presence of narrow
high-ionisation emission lines, as described in Section \ref{type2a},
with regular star-forming galaxies identified by having at least one
narrow emission line. Absorption line galaxies are objects with no
emission lines, and visible stellar absorption features such as the
Ca\,{\sc ii} H and K lines.  Stars are separated into two subgroups,
M-type stars and stars of earlier types, ranging from A to K. 

\subsection{Type 2 Seyfert Galaxies}\label{type2a}

Objects with narrow emission lines and high-ionisation line ratios are
classified as type 2 Seyfert galaxies.  The criteria used to identify
candidates are taken directly from Zakamska et al. (2003), who studied
a large number of objects extracted from SDSS DR1, except that the
only emission lines used in this work are H$\beta$, [O\,{\sc
  iii}]$\lambda$5008, H$\alpha$, and [N\,{\sc ii}]$\lambda$6583. As
the presence of the [O\,{\sc iii}]$\lambda$5008 line in the spectra is
critical for this analysis, the sample is restricted to z $<0.76$. The
selection criteria require the rest frame equivalent width (EW) of the
[O\,{\sc iii}]$\lambda$5008 to be less than $-4$\AA, the FWHM of the
([O\,{\sc iii}]) to be greater than $400$ km~s$^{-1}$, and the FWHM
for all emission lines present in each spectrum to be less than $2000$
km~s$^{-1}$.  For spectra with all four lines present, the following
must be true for selection:

\begin{equation}\label{eq:1}
\mathrm{log}\left(\frac{\mathrm{[OIII]\lambda5008}}{\mathrm{H\beta}}\right)>\frac{0.61}{\mathrm{log([NII]/H\alpha)}-0.47}+1.19
\end{equation}

\noindent For spectra at redshifts z $>0.33$, both H$\alpha$ and
[N\,{\sc ii}]$\lambda$6583 have redshifted outside the usable spectral range,
leaving only H$\beta$ and [O\,{\sc iii}]$\lambda$5008. For these cases, the
following is applied:

\begin{equation}\label{eq:2}
\mathrm{log}\left(\frac{\mathrm{[OIII]\lambda5008}}{\mathrm{H\beta}}\right)>0.3
\end{equation}

\noindent Measurements were made on the flux-calibrated spectra.
There are 96 objects that satisfy either equation \ref{eq:1} or
\ref{eq:2}, and 43 at z $<0.33$. For objects with H$\beta$ emission
lines that are too weak to be measured, a flux limit is imposed so
that the spectra may still be used in the analysis. The flux limit is
set to be 2.5 times the standard deviation calculated from a small
section of continuum at the wavelength of the absent line.

\begin{figure}
\resizebox{\hsize}{!}{\includegraphics{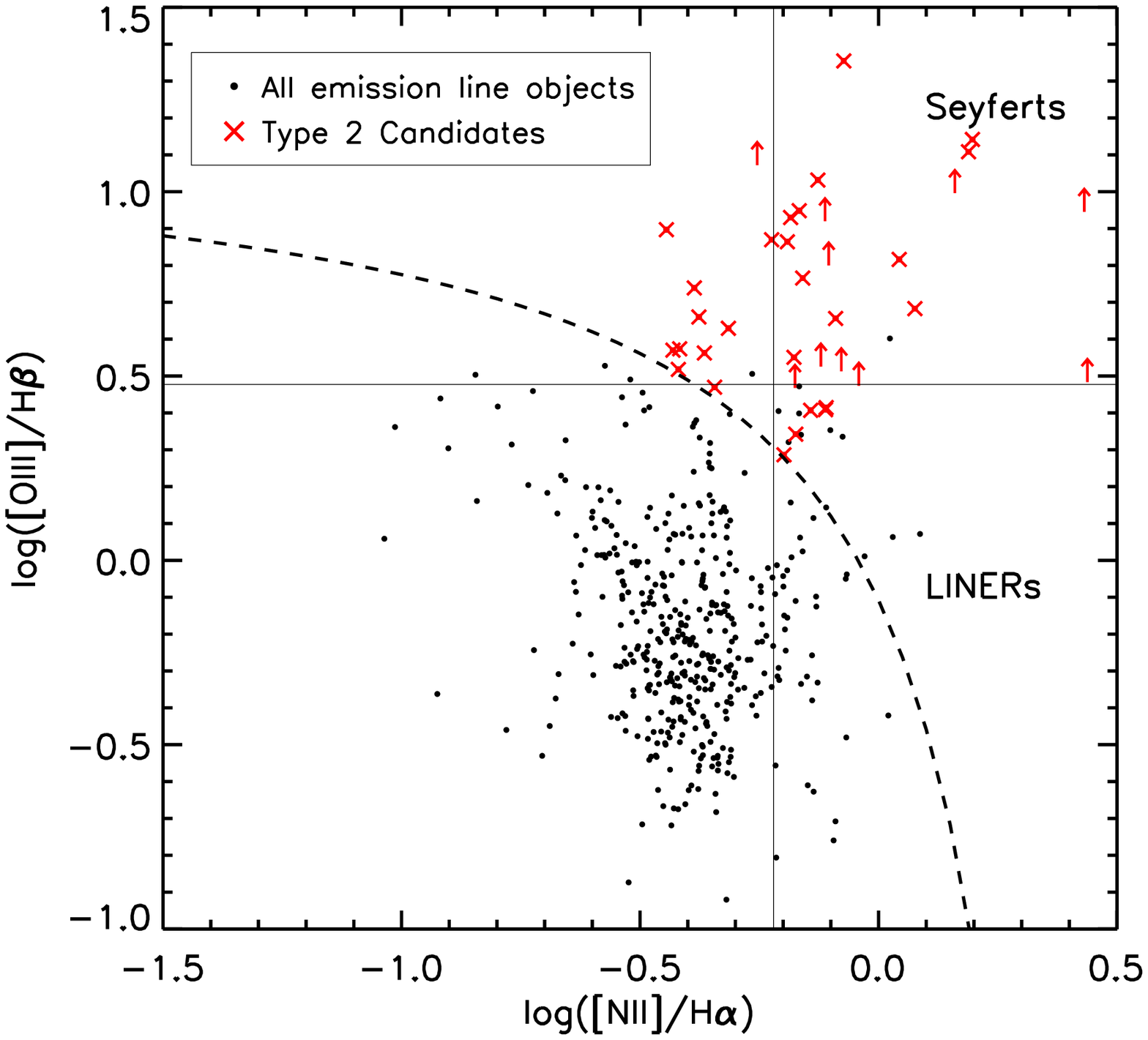}}
\caption{A BPT diagram containing all galaxies for which each of the
  four emission lines was present, excluding objects with emission
  lines broader than $1500$ km s$^{-1}$. Arrows represent objects for
  which only an upper limit to the H$\beta$ emission line flux exists.
  The dashed line is the theoretical demarcation between star-forming
  galaxies and AGN, while the solid lines divide the plot into regions
  dominated by Seyfert galaxies in the top right corner, and LINERs in
  the bottom right corner.}
\label{fig:bpt}
\end{figure}

Fig. \ref{fig:bpt} shows a standard diagnostic tool used in AGN
studies, first demonstrated by Baldwin, Phillips \& Terlevich
(hereinafter BPT, 1981). The BPT diagram separates normal star-forming
galaxies from AGN by considering emission line ratios. The dashed
demarcation line is from Kewley et al. (2001), defined by equation
\ref{eq:1}, with objects above this line having line ratios that
cannot be produced by normal star formation. Red crosses
indicate objects that have measured emission line properties
consistent with the selection criteria described above. Red arrows are
objects with only upper limits on their H$\beta$ flux. The few black
dots located above the dashed line have emission line ratios
consistent with the AGN criteria, but the measured values of the
[O\,{\sc iii}]$\lambda$5008 line equivalent width and FWHM do not meet
the specified criteria.

As the cutoff specifying broad emission lines is set to be
FWHM$>1500$~km~s$^{-1}$ (which is slightly larger than the SDSS limit
of $>1000$~km~s$^{-1}$) and the narrow line selection requirement is
FWHM$<2000$~km~s$^{-1}$, consistent with the analysis from Zakamska et
al. (2003), there is the possibility for some objects to be classified
as both broad and narrow lined with linewidths $1500<$ FWHM
$<2000$~km~s$^{-1}$. To remedy this situation, the broad line
selection was done first, followed by the narrow line selection, with
the stipulation that an object classified as broad lined cannot be
reclassified as narrow lined.

The objects identified as type 2 Seyferts lie in the same region of
the $gJK$ plot as ordinary emission line galaxies, separate from the
model quasar locus. Additional information, such as a measure
of central concentration, as described in Section \ref{SF}, would be
required for an experiment aimed specifically at their selection.
However, as type 2 objects were not the primary target of this study,
no attempt was made to select them specifically.

Much information regarding the type 2 population may be gleaned from
this sample. For the 43 objects at z $<0.33$, both H$\alpha$ and
H$\beta$ are visible in the spectra, and a measure of the dust
reddening can be estimated using the ratio of the attenuated to
observed Balmer line ratio. The distribution of E($B-V$) values can
then be determined, and applied to all 96 objects to estimate their
unreddened magnitudes. However, due to the fact that these spectra are
not of high enough quality to accurately subtract the stellar
continuum to fully reveal the emission lines, and the PSF magnitudes
show signs of host galaxy light contamination, further analysis on
this type 2 population is deferred to a later date.


\subsection{Star-forming Galaxies}\label{SF}

\begin{figure}
\resizebox{\hsize}{!}{\includegraphics{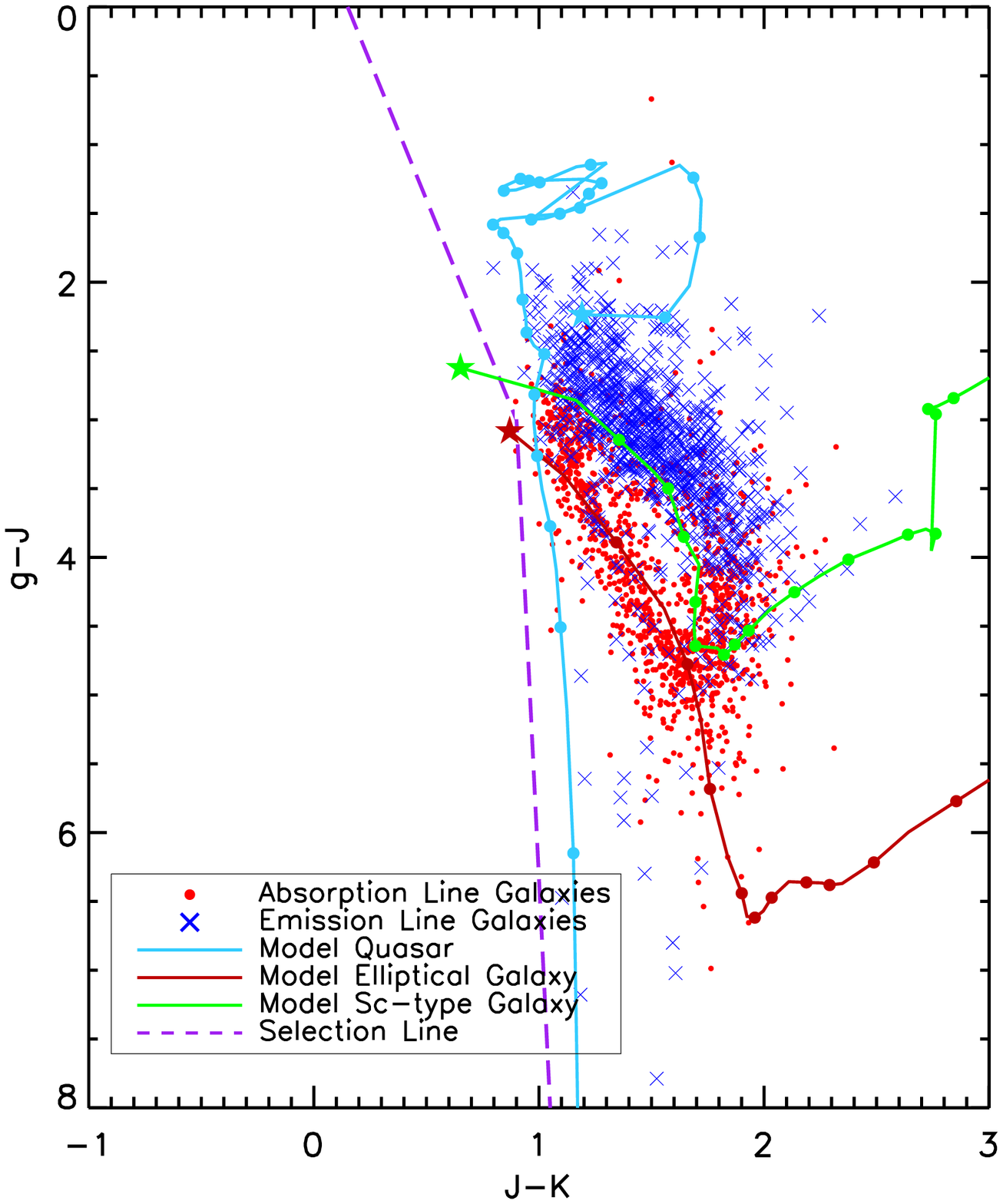}}
\caption{Location of spectroscopically confirmed galaxies on the $gJK$
  diagram. Blue crosses show emission line galaxies, with red dots
  marking absorption line galaxies. The purple selection line, blue model
  quasar and red elliptical galaxy tracks are as in Fig.
  \ref{fig:gJK1}, with the green track following the colours for a
  model Sc-type galaxy.}
\label{fig:gals}
\end{figure}

Star-forming galaxies are selected as having at least one visible
emission line, but do not satisfy the criteria for type 2 AGN as
described in the previous section. Fig. \ref{fig:bpt} shows their
location on the BPT plot as black dots below the dashed demarcation
curve. Fig. \ref{fig:gals} indicates where they lie in the $gJK$
diagram, marked as blue crosses. They are generally bluer than the
absorption line galaxies in $g-J$, but still significantly redder than
quasars at the same low redshifts. The model track for an Sc-type
galaxy is overlaid on the $gJK$ diagram as well, as in Fig.
\ref{fig:noid}. The star-forming galaxies cluster around the model
track as expected.

Star-forming galaxies are a significant contaminant in the
$gJK$-selection of quasars, as they can be morphologically compact and
lie in the region of colour space occupied by reddened quasars at $1<$
z $<3$.  However, their images tend to be much less centrally
concentrated than quasars and AGN, so imposing a restriction on a
measurement of concentration (such as the difference between the PSF
and Petrosian magnitudes from SDSS photometry, or between the
\texttt{aperMag1} and \texttt{aperMag3} measures from WFCAM
photometry) could significantly reduce the number of star-forming
galaxies considered as candidates.  This additional information could
also assist in selection of type 2 Seyfert galaxies, as they tend to
be more concentrated than star-forming galaxies as well, albeit to a
lesser degree than the quasars.

\subsection{Absorption Line Galaxies}

Aside from the stellar locus, the most densely populated region of the
$gJK$ two-colour diagram is occupied by low redshift absorption line
galaxies, shown as red dots in Fig. \ref{fig:gals}. As we did not want
to completely eliminate morphologically extended objects, but did not
have enough spectroscopic fibres available to observe them all, the
galaxy region was sparsely sampled, resulting in the hole seen in Fig.
\ref{fig:gals} at $g-J=4$, $J-K=1.5$. Most of the objects with
peculiar colours, such as the absorption line galaxies at $g-J \sim
1$, have close neighbours, or have poor SDSS photometry.


\subsection{Stars}

\begin{figure}
\resizebox{\hsize}{!}{\includegraphics{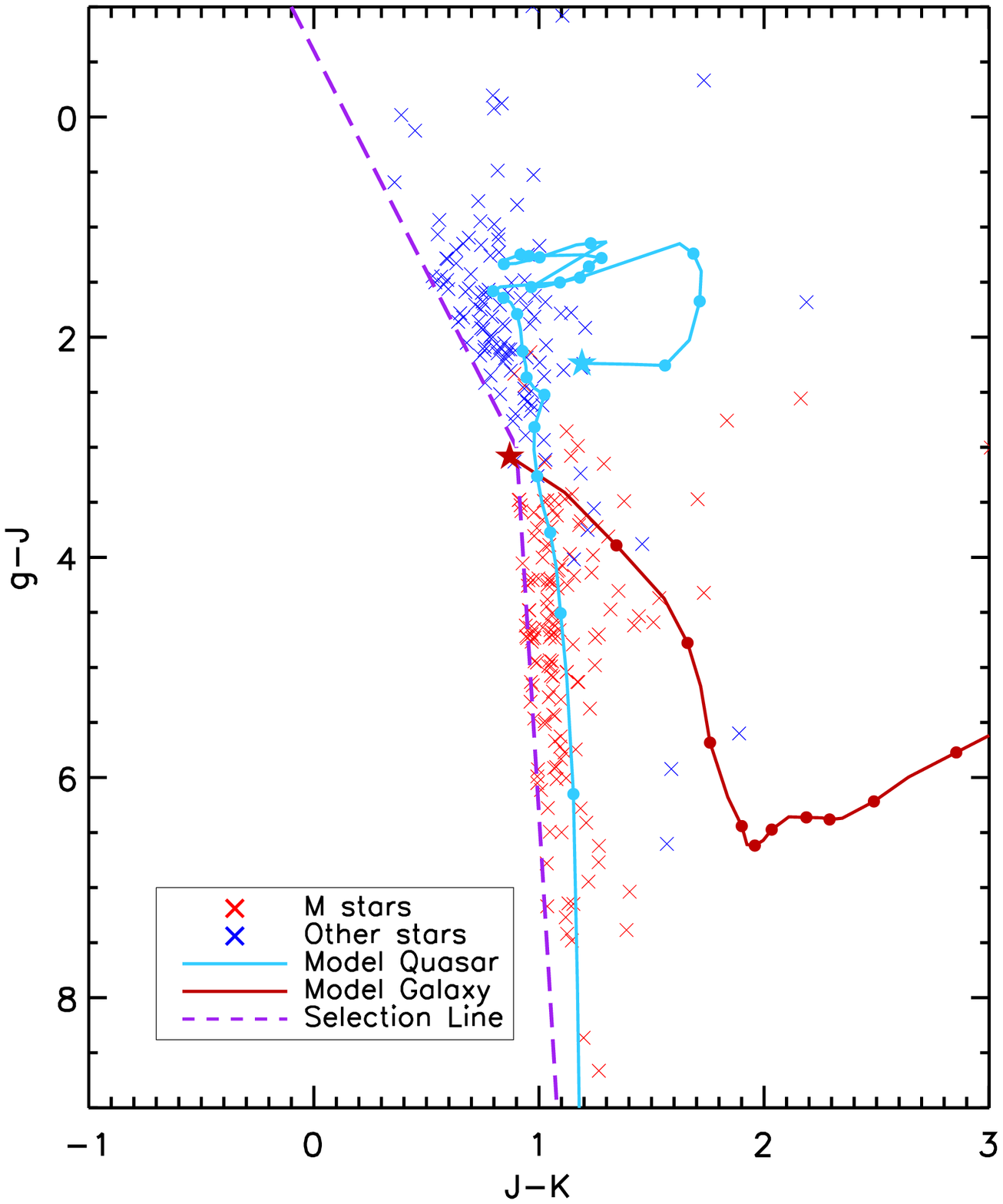}}
\caption{Location of spectroscopically confirmed stars on the $gJK$
  diagram. The majority of stars cluster near the selection line, but
  a significant number are much redder in $J-K$. Visual inspection
  reveals that most of these objects are close pairs of stars,
  resulting in atypical colours. The purple selection line, blue model
  quasar and red elliptical galaxy tracks are as in Fig.
  \ref{fig:gJK1}. Note that the y axis limits have been expanded
  slightly to include objects with extreme $g-J$ colours.}
\label{fig:stars}
\end{figure}

Stars account for nearly 10 per cent of the total number of objects
observed. Fig. \ref{fig:stars} indicates where they lie in $gJK$
colour space, with the majority clustered near the selection line. As
can be seen, there are two distinct populations, with M-type stars
redder in $g-J$ than the other types, which includes K through A-type
stars.

Stars with particularly red colours in $J-K$ are predominantly close
pairs of stars. Occasionally the objects that are red in $J-K$ are
located in the halos of bright stars, and weren't excluded by the sky
variance cut described in Section \ref{cuts}. Some objects clearly
exhibit spectral features of two stellar types, with the final
designation being determined by the type contributing most to the flux
in the AAOmega spectra. These account for most of the objects that are
classified as one type but appear in the colour space occupied by the
other type.

If reduction of stellar contamination is a priority when designing
future observations based on the $gJK$ colours, the selection line
could be moved redward in $J-K$, at the expense of rare, high redshift
objects seen in Fig. \ref{fig:gJK2}. The number of pairs of stars
observed spectroscopically that have anomalously red $J-K$ colours
could be significantly reduced by visual inspection of their images.

\end{document}

%% file: table5.tex
\begin{table*}
\caption{Photometric and spectroscopic information for each observed object.}
\label{tab:allobs}
\centering
\begin{scriptsize}
\begin{tabular}{cccccccccccc}\\ \hline
Name & UKIDSS & $g$ & $\sigma_g$ & $J$ & $\sigma_J$ & $K$ & $\sigma_K$ & E($B-V$) & Type & Redshift & SNR \\
 & Class & & & & & & & & & & \\ \hline
ULAS J121737.39-001212.2 &  1 &  21.600 &   0.049 &  17.736 &   0.051 &  16.555 &   0.050 &   0.022 &      Abs &   0.207 &     6.0  \\
ULAS J121738.61+002016.6 & -2 &  22.212 &   0.075 &  18.365 &   0.104 &  16.715 &   0.065 &   0.027 &      Abs &   0.383 &     4.2  \\
ULAS J121739.55-000124.5 &  1 &  19.328 &   0.020 &  16.334 &   0.015 &  15.210 &   0.015 &   0.026 &      Abs &   0.117 &    11.7  \\
ULAS J121741.65+002310.2 &  1 &  21.417 &   0.043 &  17.383 &   0.043 &  16.037 &   0.035 &   0.027 &      Abs &   0.257 &     5.5  \\
ULAS J121744.02+002034.8 &  1 &  21.415 &   0.042 &  18.152 &   0.086 &  16.621 &   0.060 &   0.027 &       Em &   0.373 &     7.6  \\
ULAS J121745.21+001600.5 & -2 &  20.197 &   0.025 &  17.579 &   0.051 &  16.564 &   0.057 &   0.027 &    Kstar &   0.000 &    11.0  \\
ULAS J121749.22-000143.3 & -2 &  20.780 &   0.030 &  18.260 &   0.082 &  16.657 &   0.055 &   0.027 &       Em &   0.303 &     9.4  \\
ULAS J121749.26-001129.5 & -3 &  24.263 &   0.407 &  18.678 &   0.120 &  16.922 &   0.070 &   0.023 &      Abs &   0.786 &     1.9  \\
ULAS J121749.58-000514.4 &  1 &  21.635 &   0.050 &  18.060 &   0.069 &  16.463 &   0.046 &   0.025 &    Type2 &   0.326 &     7.6  \\
ULAS J121749.87+001721.8 &  1 &  22.340 &   0.083 &  17.771 &   0.061 &  16.106 &   0.037 &   0.027 &      Abs &   0.349 &     7.1  \\
\hline\\
\end{tabular}\\
Note: The full table of 3154 objects is published in the electronic version of the paper. A portion is shown here for guidance regarding its form and content.
\end{scriptsize}
\end{table*}